\documentclass[aps,pra,reprint,onecolumn,nofootinbib,floatfix,superscriptaddress]{revtex4-2}
\usepackage[utf8]{inputenc}
\usepackage[T1]{fontenc}
\usepackage{lmodern}
\usepackage{amsmath,amssymb,bm}
\usepackage{graphicx}
\usepackage{booktabs}
\usepackage{hyperref}
\newcommand{\SI}[2]{#1\,#2}
\newcommand{\SIrange}[3]{#1--#2\,#3}
\newcommand{\micro}{\mathord{\mu}}
\newcommand{\ohm}{\Omega}
\usepackage{xcolor}
\usepackage{orcidlink}
\usepackage{microtype}
\usepackage{float}
\makeatletter
\renewcommand{\Dated@name}{}
\makeatother
\graphicspath{{figures/}}
\hypersetup{colorlinks=true,linkcolor=blue!45!black,citecolor=blue!45!black,urlcolor=blue!45!black,hypertexnames=false,pdftitle={Preparation-controlled relaxation in an overdamped RLC circuit: A pedagogical route to the spectral Mpemba effect},pdfsubject={Preparation-controlled spectral relaxation in deterministic, stochastic, and open-system descriptions},pdfkeywords={RLC circuit, spectral Mpemba analogue, modal relaxation, engineered noise, Ornstein-Uhlenbeck process, Fokker-Planck dynamics, Lindblad dynamics, open quantum systems, physics education}}

\newcommand{\dd}{\mathrm d}
\newcommand{\kb}{k_{\rm B}}
\newcommand{\Tr}{\operatorname{Tr}}
\newcommand{\lams}{\lambda_{\rm s}}
\newcommand{\lamf}{\lambda_{\rm f}}
\newcommand{\rs}{r_{\rm s}}
\newcommand{\rf}{r_{\rm f}}
\newcommand{\cs}{c_{\rm s}}
\newcommand{\cf}{c_{\rm f}}

\begin{document}
	
	\title{Preparation-controlled relaxation in an overdamped RLC circuit:\\ A pedagogical route to the spectral Mpemba effect}
	
	\author{Matheus H. dos Santos}
	\affiliation{Instituto de Física, Universidade Federal de Goiás, 74690-900, Goiânia, GO, Brazil}
	
	\author{Cler T. Garcez}
	\affiliation{Instituto de Física, Universidade Federal de Goiás, 74690-900, Goiânia, GO, Brazil}
	
	\author{Jeveson C. da Silva}
	\affiliation{Instituto de Física, Universidade Federal de Goiás, 74690-900, Goiânia, GO, Brazil}
	
	\author{G. D. de Moraes Neto\,\orcidlink{0000-0003-4273-8380}}
	\email{gdmneto@gmail.com}
	\affiliation{Department of Fundamental Sciences, Hainan Bielefeld University of Applied Sciences, Danzhou, Hainan 578101, China}
	
	\author{Norton G. de Almeida}
	\email{norton@ufg.br}
	\affiliation{Instituto de Física, Universidade Federal de Goiás, 74690-900, Goiânia, GO, Brazil}
	
	\date{}
	
	\begin{abstract}
		An overdamped resistor--inductor--capacitor (RLC) circuit can relax in a counterintuitive way: a state with more stored energy can fall below a lower-energy state even though both lose energy monotonically.  This spectral Mpemba analogue arises because the initial energy alone does not determine the relaxation history; the initial charge and current also determine how strongly the circuit excites its slow and fast decay modes.  We develop the mechanism from deterministic dynamics to stochastic and open-system descriptions.  For the classical circuit, we derive the modal preparation conditions and crossing criterion, show geometrically why equal-energy states can follow different relaxation histories, and test the effect experimentally by reconstructing charge, current, and energy from oscilloscope traces.  The ordering inversion is observed in deterministic relaxation and remains visible in an ensemble driven by externally injected white noise.  An equilibrium Ornstein--Uhlenbeck treatment then separates nonequilibrium information carried by the ensemble mean from information stored in its covariance: centered Gibbs preparations preserve their mean-energy ordering, whereas displaced or anisotropic Gaussian states can encode distinct slow and fast sectors.  Recasting the stochastic dynamics through a Fokker--Planck generator provides a bridge to Markovian quantum dynamics.  For a thermally damped quantum harmonic oscillator, the mean bare-energy excess has a single exponential decay factor, which forbids an energy-order inversion for arbitrary initial states with finite mean occupation.  A below-threshold parametric oscillator restores distinct slow and fast quadrature sectors and permits both displacement and covariance-controlled crossings.  The resulting Liouvillian formulation makes explicit that observed relaxation depends jointly on the spectrum of the generator, the modes populated by the preparation, and the modes visible to the chosen observable.
	\end{abstract}
	
	\maketitle

	\section{Introduction}
	
	An overdamped resistor--inductor--capacitor (RLC) circuit is one of the simplest systems used to teach relaxation.  Once the source is removed, the energy stored in the capacitor and inductor decreases as the resistor dissipates it.  It is therefore tempting to expect that, if two copies of the same circuit begin with different energies, the one with more energy should remain above the other throughout the relaxation.  That expectation fails when the initial states excite the circuit's two nonoscillatory decay modes in different proportions.  A preparation dominated by the fast mode can begin with more energy and nevertheless fall below a preparation dominated by the slow mode.  Both energies decrease monotonically; what changes is their ordering.
	
	This simple observation exposes a distinction that is often hidden in elementary treatments of damping.  The initial energy tells us how much energy is stored, but it does not tell us how that energy is distributed among the available relaxation modes.  In a second-order system, the initial charge and current together determine that modal composition.  The same idea can be visualized geometrically: states with the same energy lie on one constant-energy curve, yet different points on that curve can relax on very different time scales.  The problem therefore provides a concrete route from a familiar ordinary differential equation to state-space dynamics, eigenmodes, energy geometry, and experimentally controllable initial conditions.
	
	Preparation-dependent relaxation is closely related to the spectral interpretation of the Mpemba effect. The historical Mpemba problem asks how a state initially farther from equilibrium can overtake another during relaxation \cite{MpembaOsborne1969, BurridgeLinden2016}. Its interpretation depends on the observable and on the criterion used to compare relaxation \cite{Jeng2006, Balazovic2012, Ibekwe2016}. In linear systems, a particularly transparent mechanism emerges from the spectrum of the dynamical generator: different initial states can have different overlaps with the slowest decaying modes, so suppressing a slow contribution can accelerate the observed approach to equilibrium \cite{Lu2017, Klich2019, Teza2026}. Similar modal mechanisms and relaxation-rate inversions have been identified in nonlinearly driven granular gases and inelastic hard-disk systems \cite{Megias2022, Santos2026}, where nonequilibrium preparation and memory effects dictate the modal weight distribution. Frameworks based on thermomajorization and Liouvillian speed limits further demonstrate how the suppression of slow-decaying sectors controls thermal quenches in classical and quantum regimes \cite{Vu2025, Hayakawa2026}, while active stochastic dynamics offer concrete parallels for mode excitation in anisotropic potentials \cite{Biswas2025}. We use the term spectral Mpemba analogue for the RLC ordering inversion studied here. The terminology emphasizes the common modal mechanism without identifying the electrical experiment with the thermal freezing problem that motivated the original effect.
	
	The RLC circuit is well suited to this discussion because it connects several topics that students often encounter separately: linear differential equations, damped oscillators, electrical transients, energy storage, dissipation, phase-space representations, and laboratory measurement.  RLC systems have long served as teaching platforms for transient and resonance phenomena \cite{Faleski2006,Cafarelli2012,Mungan2022,Kowalski2024RLC,TorrienteGarcia2024,Ziese2026RLC}, while classical resonators have also been used to make more advanced spectral ideas experimentally accessible \cite{GarridoAlzar2002EIT,Harden2011DoubleEIT,Satpathy2012Fano}.  This is useful pedagogically because familiarity with circuit components does not automatically imply a coherent picture of the dynamics: studies of undergraduate circuit reasoning have documented difficulties in connecting symbolic relations, component behavior, and physical interpretation \cite{LiSingh2012CircuitElements,LiuPanZhangBao2022,Bauman2024Circuits,Gross2026BEAST,Steinmetz2026Circuits}.  Here the same circuit supports both the mathematical construction and a direct experimental test of the resulting relaxation picture.
	
	The deterministic circuit forms the central thread of the article.  We prepare states close to the slow and fast eigendirections, reconstruct charge, current, and energy from oscilloscope measurements, and observe the predicted reversal of the energy ordering.  The equal-energy construction provides a complementary view: fixing the initial energy does not fix the subsequent relaxation because the modal weights still depend on the direction of the state in phase space.  A second experimental protocol adds an externally generated white-noise drive.  The resulting ensemble retains a visible ordering inversion, while the measured energy can be separated into contributions associated with the ensemble mean and its fluctuations.  This driven-noise experiment is treated as an engineered stochastic perturbation; no effective temperature is assigned without an independent calibration of the noise reaching the RLC circuit.
	
	Thermal fluctuations provide a natural theoretical extension because the same resistor that dissipates energy also generates Johnson--Nyquist noise at nonzero temperature \cite{Johnson1928,Nyquist1928}.  In that setting a single trajectory is replaced by a probability distribution, and the deterministic state-space picture grows into the mean--covariance dynamics of an Ornstein--Uhlenbeck process \cite{Gillespie1996BrownianJohnson,Vasziova2010,Tothova2011}.  This extension separates three physically different possibilities.  Changing only the temperature of a centered Gibbs state preserves the mean-energy ordering, whereas a displaced thermal state can carry slow or fast information in its mean, and a centered anisotropic Gaussian state can carry the same information in its covariance.  The circuit therefore provides a direct bridge from elementary relaxation modes to stochastic descriptions in which nonequilibrium information can reside in different statistical moments.
	
	Section~V carries the spectral construction into classical and quantum generator descriptions.  The Ornstein--Uhlenbeck circuit is first recast in Fokker--Planck form, where the deterministic drift remains embedded in the probability generator and the first- and second-moment sectors acquire their own decay rates.  For a Markovian quantum harmonic oscillator, the Lindblad equation can likewise be written as a Wigner Fokker--Planck equation, making the common drift--diffusion structure explicit while preserving the physical distinction between a classical probability density and a quantum quasiprobability \cite{Gorini1976GKSL,Lindblad1976,Manzano2020Lindblad,Campaioli2024QME}.  The ordinary thermally damped oscillator then gives a useful negative result: its mean bare-energy excess has a single decay factor, so its energy ordering cannot reverse under that generator.  An anisotropic parametric oscillator restores distinct slow and fast quadrature sectors and permits both displacement and covariance-controlled energy crossings.  The final Liouvillian formulation separates two ingredients that already appear in the circuit problem: whether a decay mode is populated by the initial preparation and whether that mode is visible to the chosen observable.  In this form, the circuit provides a concrete starting point for a broader spectral description of relaxation without implying a thermodynamic equivalence between the classical and quantum systems.
	
	The prerequisites increase across the three parts of the article.  The deterministic analysis and its physical interpretation require standard undergraduate RLC circuit theory, linear ordinary differential equations, and basic linear algebra; reproducing the measurements additionally requires introductory laboratory experience with circuit construction and oscilloscope acquisition.  The stochastic treatment in Sec.~IV moves to a more advanced level, using probability distributions, stochastic differential equations, and Gaussian mean--covariance dynamics.  Section~V is intended primarily for beginning graduate students who already have some familiarity with density operators, Markovian master equations, and open quantum systems.  The companion notebooks contain the longer intermediate derivations, numerical checks, experimental reconstruction procedures, and reproducible calculations, including the detailed classical-to-quantum phase-space derivations used in Sec.~V.  Across these different levels, the underlying physical question remains the same: relaxation depends not only on the initial distance from equilibrium, but also on how the preparation populates the available decay modes and on which of those modes are visible in the quantity being observed.
	
	\section{Deterministic RLC relaxation: why the initial state matters}
	\label{sec:deterministic_modes}
		\begin{figure}[htbp]
		\centering
		\includegraphics[width=\textwidth]{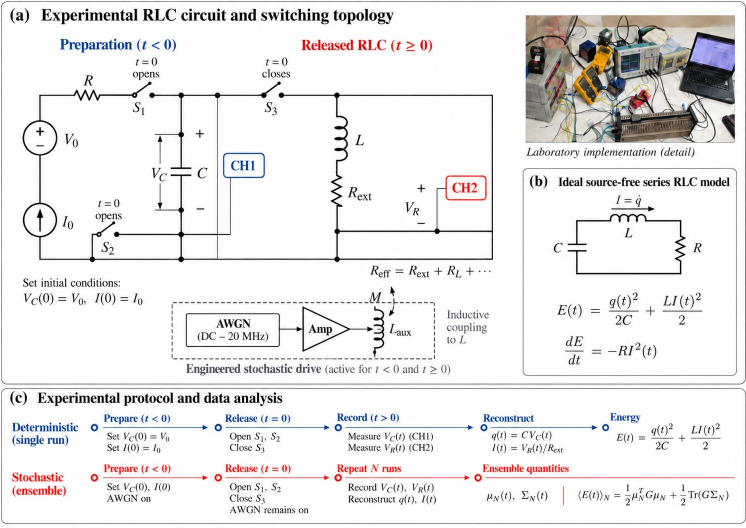}
		\caption{Experimental platform and analysis map. (a) Preparation and release circuit, measurement channels CH1 ($V_C$) and CH2 ($V_R$), current-sensing resistor $R_{\rm ext}$, and auxiliary inductive branch used for the engineered-noise drive; the inset shows the laboratory implementation. (b) Ideal source-free series RLC model used after release. (c) Deterministic and stochastic reconstruction paths. Channel, sign, and release-time conventions are given in Sec.~\ref{sec:experimental_reconstruction}.}
		\label{fig:apparatus_protocol}
	\end{figure}

	An overdamped series RLC circuit is usually introduced as a simple example of a system that returns to equilibrium without oscillating.  The familiar description, however, hides an important piece of physics: the circuit has two distinct nonoscillatory decay modes, and the initial charge and current determine how strongly each one is excited.  Two preparations of the same circuit can therefore store the same energy, or even start with different energies, and still relax in very different ways.  This makes the RLC circuit a particularly transparent setting for separating the amount of stored energy from the way that energy relaxes.  Standard treatments of RLC transients can be found in Refs.~\cite{AlexanderSadiku2017,NilssonRiedel2019,Cafarelli2012}.
	
	After the source used to prepare the circuit is disconnected, the capacitor charge $q(t)$ obeys
	\begin{equation}
		L\ddot q+R\dot q+\frac{q}{C}=0,
		\label{eq:rlc_second_order}
	\end{equation}
	with $I=\dot q$.  A second-order equation needs two initial values, which here have an immediate physical meaning: the initial charge $q(0)$ and the initial current $I(0)$.  It is therefore useful to regard $X=(q,I)^T$ as the state of the circuit and to rewrite the same dynamics as
	\begin{equation}
		\dot X=A X,
		\qquad
		A=\begin{pmatrix}
			0&1\\ -1/(LC)&-R/L
		\end{pmatrix}.
		\label{eq:A_matrix}
	\end{equation}
	Nothing new has been assumed in going from Eq.~\eqref{eq:rlc_second_order} to Eq.~\eqref{eq:A_matrix}; one second-order ODE has simply been written as two coupled first-order ODEs.\footnote{For readers who have only recently met ordinary differential equations, this first-order-system viewpoint is a standard way of treating higher-order linear ODEs.  A detailed undergraduate introduction is given in Ref.~\cite{BoyceDiPrima2017}.}
	
	The natural quantity for comparing different preparations is the energy stored in the capacitor and inductor,
	
	\begin{equation}
		E(t)=\frac{q(t)^2}{2C}+\frac{L I(t)^2}{2} 
	\end{equation}
	
	Differentiating and using the circuit equation gives the particularly simple result
	\begin{equation}
		\dot E=-R I^2\leq0.
		\label{eq:energy_dissipation}
	\end{equation}
	Thus every source-free trajectory considered here loses energy monotonically.  An inversion of the energy ordering does not require either trajectory to gain energy: two decreasing curves can cross because their initial states contain different mixtures of the slow and fast decay modes.
	
	Equation~\eqref{eq:energy_dissipation} also gives a first hint of why the initial state matters.  The resistor removes energy through the current.  Two states can have the same total energy while storing different fractions of it in the capacitor and the inductor, so their instantaneous rates of energy loss need not be the same.  This becomes easy to see if we rescale charge and current according to
	\begin{equation}
		z_1=\frac{q}{\sqrt C},
		\qquad
		z_2=\sqrt L\,I.
		\label{eq:energy_normalized_coordinates}
	\end{equation}
	In these variables $E=(z_1^2+z_2^2)/2$.  Equal-energy states therefore lie on a circle in the $(z_1,z_2)$ plane: the radius tells us how much energy is stored, while the position around the circle tells us how that energy is divided between charge and current.  Changing the direction at fixed radius can therefore change the subsequent relaxation without changing the initial energy.
	
	The two relaxation rates follow directly from the textbook solution of Eq.~\eqref{eq:rlc_second_order}.  Trying $q(t)\propto e^{\lambda t}$ gives the characteristic equation $\lambda^2+(R/L)\lambda+1/(LC)=0$.  The circuit is overdamped when $R>R_c$, where $R_c=2\sqrt{L/C}$, and the two roots are then real and negative:
	\begin{equation}
		\lams=-\frac{R}{2L}+\sqrt{\left(\frac{R}{2L}\right)^2-\frac{1}{LC}},
		\qquad
		\lamf=-\frac{R}{2L}-\sqrt{\left(\frac{R}{2L}\right)^2-\frac{1}{LC}}.
		\label{eq:eigenvalues}
	\end{equation}
	We label them so that $\lamf<\lams<0$.  The exponential $e^{\lamf t}$ disappears more rapidly and will be called the fast mode; $e^{\lams t}$ survives longer and will be called the slow mode.  In other words, the familiar two-root solution of an overdamped second-order ODE already contains two possible relaxation time scales.
	
	The same statement has a useful geometrical form.  For a single exponential contribution, $I=\dot q=\lambda q$, so each root defines a straight line in the $(q,I)$ plane.  Convenient representatives of these two directions are $\rs=(1,\lams)^T$ and $\rf=(1,\lamf)^T$.\footnote{Only the direction matters: multiplying either vector by any nonzero constant gives the same physical line in state space.  This is why ``eigendirection'' is often more intuitive here than the normalization-dependent word ``eigenvector''.  For an accessible phase-plane discussion of linear systems, see Ref.~\cite{Strogatz2015}.}  A general state is a combination of the two,
	\begin{equation}
		X(t)=\cs\rs e^{\lams t}+\cf\rf e^{\lamf t}.
		\label{eq:modal_solution}
	\end{equation}
	The coefficients $c_s$ and $c_f$ tell us how much of the slow and fast mode was placed in the initial state.  Solving for them from $q(0)$ and $I(0)$ gives
	\begin{equation}
		\cs=\frac{I(0)-\lamf q(0)}{\lams-\lamf},
		\qquad
		\cf=\frac{\lams q(0)-I(0)}{\lams-\lamf}.
		\label{eq:modal_coefficients}
	\end{equation}
	The practical consequence is immediate.  If $I(0)=\lamf q(0)$, then $c_s=0$ and the slow mode is absent: the state follows only the fast decay.  If $I(0)=\lams q(0)$, then $c_f=0$ and the state follows only the slow decay.  The modal decomposition therefore gives a direct preparation rule: by setting the initial charge and current in the appropriate ratio, the same physical circuit can be made to relax predominantly on one of two very different time scales.
	
	There are two time conventions worth keeping separate.  A modal amplitude decays with time constant $1/|\lambda_j|$, while the energy of a pure mode is quadratic in that amplitude and therefore decays with time constant $1/(2|\lambda_j|)$.\footnote{For a pure mode, $q$ and $I$ both carry a factor $e^{\lambda_j t}$, so the quadratic energy carries $e^{2\lambda_j t}$.  This distinction matters when a voltage or current trace is compared with an energy trace.}  We denote the latter by $\tau_{E,j}=1/(2|\lambda_j|)$.
	
	Figure~\ref{fig:apparatus_protocol} connects this preparation picture to the laboratory system and summarizes how the deterministic and stochastic analysis branches are constructed from the measured voltages.

	The separation of the two time scales is conveniently measured by a dimensionless ratio.  The conventional damping parameter is $\alpha=R/(2L\omega_0)$ with $\omega_0=1/\sqrt{LC}$, and the overdamped regime corresponds to $\alpha>1$, so that the circuit equation takes the canonical form of a damped harmonic oscillator:
	
	\begin{equation}
		\ddot q + 2\alpha \omega_0 \dot q + \omega_0^2 q =0.    
	\end{equation}
	For comparing slow and fast preparations directly, it is more transparent to use the ratio $\rho=|\lamf/\lams|>1$, which is also equal to $\tau_{E,s}/\tau_{E,f}$ for the pure-mode energy time scales.  Thus $\rho=10$ simply means that the slow-mode energy persists ten times longer than the fast-mode energy.  The two parameterizations are equivalent: $\alpha=(\sqrt\rho+1/\sqrt\rho)/2$.  For orientation, $\alpha=1.50$ corresponds to $\rho\simeq6.85$, while $\alpha=1.75$ corresponds to $\rho\simeq10.15$.
	
	This rate ratio can be used in reverse as a circuit-design variable.  If an inductance $L$ is already available and we choose both $\rho$ and the desired slow energy time $\tau_{E,s}$, the resistance and capacitance follow from
	\begin{equation}
		R=\frac{L(1+\rho)}{2\tau_{E,s}},
		\qquad
		C=\frac{4\tau_{E,s}^2}{L\rho}.
		\label{eq:inverse_design_tau}
	\end{equation}
	This relation reverses the usual circuit problem: instead of asking which decay rates follow from a chosen set of components, one can choose the desired time-scale separation first and determine suitable component values.  There is no universal optimum for $\rho$.  If the modes are too close, their different relaxation histories separate only slowly; if the separation is extremely large, the fast transient may fall below the available time resolution while the slow tail remains long.  For the parameter ranges considered here, $\rho$ of order $5$--$15$ gives an easily resolved separation.  As one concrete design point, $\rho=10$, $L=\SI{100}{mH}$, and $C=\SI{100}{\micro F}$ give $R\simeq\SI{110}{\ohm}$, $\lams\simeq-\SI{100}{s^{-1}}$, and $\lamf\simeq-\SI{1000}{s^{-1}}$.  For the laboratory component values, $R_c\simeq\SI{19.8}{\ohm}$ and $R_{\rm comp}=\SI{27.9}{\ohm}$, so the circuit is safely overdamped; the larger data-derived diagnostic $R_{\rm dyn}$ changes the inferred rate separation but not that classification.
	
	Consider one state prepared on the fast direction and another on the slow direction.  Their energies have the simple forms
	\[
	E_f(t)=E_f(0)e^{2\lamf t},
	\qquad
	E_s(t)=E_s(0)e^{2\lams t}.
	\]
	Suppose that the fast state begins with more energy, $E_f(0)>E_s(0)$.  Because $|\lamf|>|\lams|$, its energy nevertheless falls more rapidly, and the two curves cross at
	\begin{equation}
		t_\times=\frac{\ln[E_s(0)/E_f(0)]}{2(\lamf-\lams)}>0.
		\label{eq:crossing_time}
	\end{equation}
	This is a deterministic inversion of the energy ordering.  Both energies decrease throughout the process; the initially higher-energy state falls below the initially lower-energy state because it was prepared almost entirely in the faster relaxation mode.  In terms of $\rho$ and the slow-mode energy time, the same result becomes $t_\times/\tau_{E,s}=\ln[E_f(0)/E_s(0)]/(\rho-1)$.  A larger initial energy ratio delays the crossing, whereas a larger separation between the two decay rates brings it forward.
	
	For the reference values $\lams=-\SI{100}{s^{-1}}$ and $\lamf=-\SI{1000}{s^{-1}}$, choosing $E_s(0)=\SI{1}{\micro J}$ and $E_f(0)=\SI{10}{\micro J}$ gives $t_\times=\SI{1.279}{ms}$ and $E_\times\simeq\SI{0.774}{\micro J}$.  The crossing therefore occurs on the millisecond scale and can be checked directly from the two exponentials.  The more general case in which both modes are present is algebraically richer because the physical energy is quadratic in $q$ and $I$.  Substituting Eq.~\eqref{eq:modal_solution} into the energy produces contributions with decay rates $2\lams$, $\lams+\lamf$, and $2\lamf$.  The cross term matters for a mixed preparation, whereas the pure-mode inversion already follows from the different decay rates alone.
	
\begin{figure}[htbp]
		\centering
		\includegraphics[width=\textwidth]{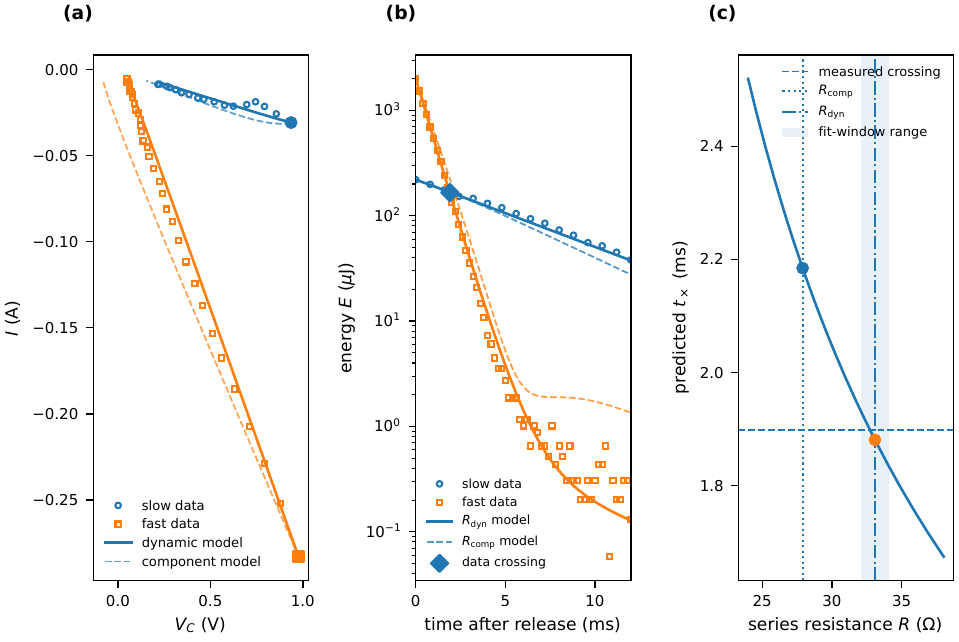}
		\caption{Deterministic preparation-controlled relaxation. (a) Reconstructed phase-space trajectories compared with component-based and dynamically inferred models. (b) Reconstructed energies and measured crossing (diamond). (c) Predicted crossing time versus series resistance, showing $R_{\rm comp}$, $R_{\rm dyn}$, the measured crossing, and the tested fit-window range. The shaded interval denotes fit-window sensitivity, not a statistical confidence interval or an independent resistance calibration.}
		\label{fig:deterministic_data}
	\end{figure}

	The laboratory data provide the direct version of this comparison in
	Fig.~\ref{fig:deterministic_data}.  Panel (a) shows that the measured slow and
	fast preparations follow clearly different paths in the reconstructed
	\((V_C,I)\) plane, with the dynamical model tracking the observed trajectories
	more closely than the component-only prediction.  Panel (b) then displays the
	same distinction in the energy itself: the fast state begins with substantially
	more stored energy, yet its more rapid modal decay drives it below the slow
	state at a finite time while both curves remain decreasing.  Panel (c) makes
	the quantitative sensitivity transparent.  Because the two modal rates depend
	on the total damping, even a modest change in the effective series resistance
	shifts the predicted crossing appreciably; the spread shown there records the
	dependence of the dynamical diagnostic on the fitting window rather than a
	statistical confidence interval.
	
	The same geometry also allows a more stringent comparison: an entire family of preparations can be chosen with exactly the same initial energy $E_0$.  In the physical variables such a family can be written as
	\begin{equation}
		q_0(\theta)=\sqrt{2CE_0}\cos\theta,
		\qquad
		I_0(\theta)=\sqrt{\frac{2E_0}{L}}\sin\theta.
		\label{eq:ellipse_param}
	\end{equation}
	In the energy-normalized plane these states are simply points around one circle.  Moving around the circle leaves the radius, and hence the initial energy, unchanged, but it continuously changes the coefficients $c_s$ and $c_f$ in Eq.~\eqref{eq:modal_coefficients}.  The pure fast and pure slow states occur at angles satisfying $\tan\theta_f=\lamf\sqrt{LC}$ and $\tan\theta_s=\lams\sqrt{LC}$; each direction has an antipodal point at $\theta+\pi$, corresponding to the same mode with the opposite overall sign.
	
	This gives a particularly clear way to separate energy from relaxation history.  Every point on the circle begins with the same energy, but points with a noticeable slow-mode component retain a long tail, while the pure-fast direction has no slow component at all.  A convenient way to compare this family at a fixed observation time $t_*$ is through the retained-energy fraction $\mathcal R(t_*;\theta)=E(t_*;\theta)/E_0$.  This fraction changes around the circle even though every preparation has the same $E_0$.  Equal initial energy therefore does not imply equal relaxation.

	The equal-energy construction also makes the sensitivity of the fast preparation easy to understand.  Suppose we aim for $I_0=\lamf q_0$ but miss the required current by a small amount $\delta I$.  Then
	\[
	I_0=\lamf q_0+\delta I,
	\qquad
	c_s=\frac{\delta I}{\lams-\lamf}.
	\]
	The unwanted slow-mode amplitude is therefore linear in the preparation error.  Its contribution to the late-time energy is quadratic in $\delta I$ and decays only as $e^{2\lams t}$, so a component that is initially very small can become the visible tail after the fast mode has disappeared.  This is why preparation precision and observation time cannot be considered independently: the longer we watch the nominally fast state, the more sensitive we become to a small residual slow component.
	
	The deterministic picture is controlled by two independent features of the initial state.  The radius in the energy-normalized plane fixes the initial energy, while the direction fixes the mixture of slow and fast modes.  A preparation near the fast direction suppresses the long-lived component; a preparation near the slow direction enhances it.  This is why a higher-energy fast state can cross below a lower-energy slow state, and why equal-energy states can still follow very different relaxation histories.  The measured circuit provides a direct test of these preparation-dependent predictions.
	
	\section{Laboratory realization and energy reconstruction}
	\label{sec:experiment}
	\label{sec:experimental_apparatus}
	
	A central motivation of our experimental design was to provide a transparent and accessible setup suitable for advanced undergraduate laboratories. Rather than relying on specialized high-frequency instrumentation, complex switching topologies, or expensive thermal bath chambers, the apparatus employs standard off-the-shelf laboratory components, basic oscilloscope acquisitions, and low-cost relays. This minimal-resource implementation makes the physical mechanism underlying the ordering inversion directly observable from raw voltage traces, emphasizing how fundamental state-space concepts, modal dynamics, and energy geometry can be effectively demonstrated in an educational setting without sacrificing quantitative rigor.
	
	The preparation dependence described in Sec.~\ref{sec:deterministic_modes} can be tested directly in a series RLC circuit because both state variables are experimentally accessible. The capacitor voltage determines the charge, while the voltage across a known series resistor determines the current. The same measurements therefore provide the instantaneous electrical energy without differentiating a noisy voltage trace.
	
	The circuit used an inductance $L=\SI{44.4}{mH}$ and a capacitance $C=\SI{454}{\micro F}$, with quoted component uncertainties of $\SI{1.2}{mH}$ and $\SI{8}{\micro F}$, respectively. Current was measured through an external sensing resistor $R_{\rm ext}=\SI{22.2(7)}{\ohm}$. The component-level series resistance was $R_{\rm comp}=\SI{27.9(10)}{\ohm}$, while the inductor winding contributed $\SI{5.7(5)}{\ohm}$. At the central values, $R_{\rm ext}$ plus the winding resistance is consistent with $R_{\rm comp}$. The distinction between these resistances is important: $R_{\rm ext}$ converts the measured resistor voltage into current, whereas the total series resistance controls the decay rates of the RLC dynamics.
	
	Figure~\ref{fig:apparatus_protocol} summarizes the preparation and measurement arrangement, the ideal post-release model, and the deterministic and stochastic analysis paths used in this section. The capacitor voltage $V_C$ is recorded on CH1 and the voltage $V_R$ across the sensing resistor on CH2. Before release, a DC source and an auxiliary pre-drive establish the desired capacitor charge and current. A high-speed relay then isolates the preparation circuit, leaving the series RLC loop to relax freely. The fast and slow deterministic preparations were chosen to lie close to the modal conditions derived in Sec.~\ref{sec:deterministic_modes}: $I(0)\simeq\lambda_f C V_C(0)$ for the fast state and $I(0)\simeq\lambda_s C V_C(0)$ for the slow state. The fast preparation was given the larger initial energy so that the predicted ordering inversion could be observed directly.
	
	\label{sec:experimental_reconstruction}
	The preparation sources used to establish the initial capacitor
	voltage $V_C(0)$ and inductor current $I_L(0)$ were disconnected by
	a programmed trigger, initiating the source-free RLC relaxation. The
	effective onset of the deterministic relaxation was subsequently
	identified by visual inspection of the measured voltage and current
	decays at approximately $t_{\rm raw}=7.2\,\mathrm{ms}$. This common
	onset was used to define
	\begin{equation}
		t_{\rm rel}
		=
		t_{\rm raw}-7.2\,\mathrm{ms},
		\label{eq:release_time}
	\end{equation}
	with the same time translation applied to the fast and slow
	deterministic records. The two traces were not shifted independently
	or aligned by fitting their crossing positions. Because the transient onset was identified by inspecting the initial voltage drop, $t_{\text{raw}} \simeq 7.2\,\text{ms}$ establishes a common experimental time origin across measurements rather than an independently calibrated hardware timestamp. Applying the identical time translation $t_{\text{rel}} = t_{\text{raw}} - 7.2\,\text{ms}$ to both fast and slow traces ensures that uniform shifts in the chosen origin leave the relative time interval until energy crossing, $t_\times$, invariant.

	At each sample, charge, current, and energy are reconstructed from the two measured voltages:
	\begin{equation}
		q(t)=C V_C(t),
		\qquad
		I(t)=-\frac{V_R(t)}{R_{\rm ext}},
		\qquad
		E_{\rm exp}(t)=\frac{C V_C^2(t)}{2}
		+\frac{L I^2(t)}{2}.
		\label{eq:experimental_energy_reconstruction}
	\end{equation}
	The minus sign in the current definition follows the probe polarity.\footnote{Changing this sign would reflect the trajectory across the $q$ axis in phase space but would leave the reconstructed energy unchanged because the inductive contribution depends on $I^2$.} Since both $V_C$ and $V_R$ are measured directly, the current does not have to be estimated by numerically differentiating $V_C$. The raw channels are used without smoothing, filtering, normalization, or baseline subtraction.
	
	For two deterministic traces, the energy crossing is identified by the first post-release change of sign of $E_f(t)-E_s(t)$. Linear interpolation between the two samples that bracket this sign change gives the crossing time. In the present deterministic data, the fast preparation begins at approximately $\SI{1.993}{mJ}$ and the slow preparation at $\SI{0.220}{mJ}$. Their reconstructed energies cross at approximately
	$t_{\times}^{\rm det}\simeq1.9\,\mathrm{ms}$, where
	$E_{\times}^{\rm det}\simeq166\,\mu\mathrm{J}$. The two curves remain individually decreasing, as expected from $\dot E=-RI^2$, but their ordering reverses because the initially higher-energy state was prepared much closer to the fast relaxation direction.
	
	Figure~\ref{fig:deterministic_data} brings the state-space and energy views of the deterministic experiment together. In panel (a), the reconstructed slow and fast trajectories are compared with the component-based and dynamically inferred models; panel (b) shows the corresponding energy-order inversion and measured crossing. Panel (c) isolates the main quantitative sensitivity in the comparison: using the component value $R_{\rm comp}=\SI{27.9}{\ohm}$ gives a predicted crossing near $\SI{2.18}{ms}$, whereas a fit-window analysis of the transients gives a median dynamical diagnostic $R_{\rm dyn}\simeq\SI{33.08}{\ohm}$. The central 16--84\% spread across the tested windows is $[32.49 -33.52]{\ohm}$, and the full tested span is $[32.07-34.10]{\ohm}$. The shaded range in Fig.~\ref{fig:deterministic_data}(c) therefore represents fit-window sensitivity rather than a confidence interval or a replacement component calibration. The shift from $R_{\rm comp}$ shows that relatively small additional losses from switches, contacts, winding details, or other series elements can noticeably change the two decay rates and hence the predicted crossing time.
	
	The deterministic records contain 2500 samples per trace, with
	sampling intervals of approximately $40\,\mu\mathrm{s}$ for the fast
	trace and $400\,\mu\mathrm{s}$ for the slow trace. Because the
	release time was identified visually, the value
	$t_{\rm raw}\simeq7.2\,\mathrm{ms}$ should be understood as an
	experimental time-origin estimate rather than as an independently
	calibrated switching delay. The engineered-noise data are retained
	on their trigger-time coordinate, so their absolute crossing time is
	treated separately from the release-aligned deterministic value.
	
	\label{sec:experimental_noise}
	The noisy data set contains twenty traces, ten for the fast preparation and ten for the slow preparation, with a sampling interval of approximately $\SI{100}{\micro s}$. Each run is first converted to an energy trace using Eq.~\eqref{eq:experimental_energy_reconstruction}; only then are the energies averaged. This order matters because energy is quadratic in voltage and current.\footnote{In general, the energy calculated from the mean voltage and mean current is not equal to the mean of the individual energies. Fluctuations contribute through the second moments, which is precisely the contribution isolated below by the covariance term.}
	
\begin{figure}[htbp]
		\centering
		\includegraphics[width=\textwidth]{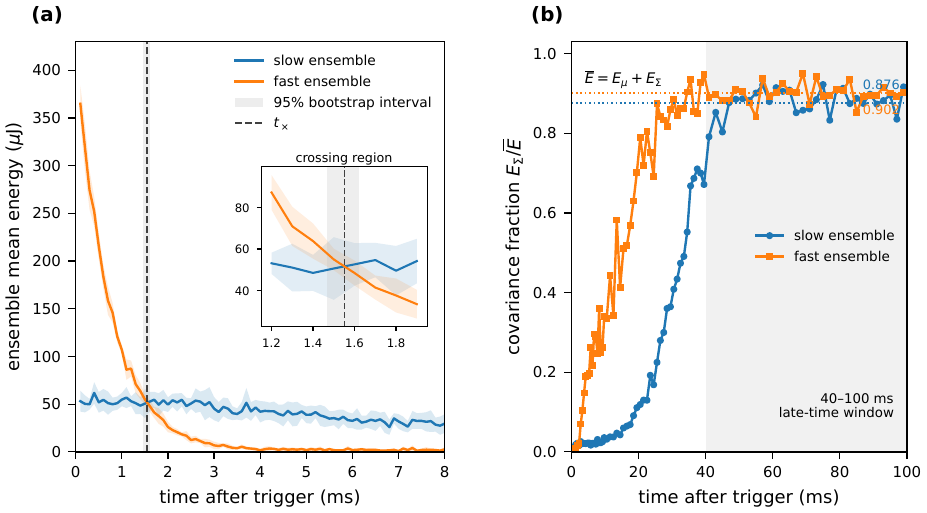}
		\caption{Engineered-noise relaxation and moment decomposition. (a) Ensemble-mean energies of the slow and fast preparations on the trigger-time coordinate. The colored bands show the run-to-run standard deviations, the dashed line marks the measured crossing $t_\times^{\rm noise}=\SI{1.55}{ms}$, and the gray band gives the 95\% trajectory-bootstrap interval $[1.47-1.62]{ms}$. The inset enlarges the crossing region. (b) Fraction of the ensemble energy carried by the covariance contribution, $E_\Sigma/\overline E$, for the same two preparations. The shaded region marks the $40$--$100$ ms late-time window; the dotted levels indicate the corresponding averages, approximately $0.902$ for the fast ensemble and $0.876$ for the slow ensemble. The drive is externally engineered AWGN rather than a calibrated equilibrium thermal bath.}
		\label{fig:stochastic_data}
	\end{figure}

	For preparation $j$, let $E_{j,r}(t)$ denote the energy in run $r$. The plotted mean and run-to-run standard deviation are
	\[
	\overline E_j(t)=\frac{1}{N}\sum_{r=1}^{N}E_{j,r}(t),
	\qquad
	s_j(t)=
	\left[
	\frac{1}{N-1}\sum_{r=1}^{N}
	\bigl(E_{j,r}(t)-\overline E_j(t)\bigr)^2
	\right]^{1/2},
	\quad N=10.
	\]
	The bands in Fig.~\ref{fig:stochastic_data} show $\overline E_j\pm s_j$. They describe the spread of the measured trajectories rather than the uncertainty of the mean.\footnote{The standard deviation answers how widely individual runs are distributed. The standard error of the mean would answer a different question: how precisely the ensemble mean has been estimated from a finite sample.}
	
	The energy of the noisy ensemble can also be separated into the part carried by the mean state and the part carried by fluctuations around that mean. If $\mu_j$ and $\Sigma_j$ are the sample mean vector and covariance matrix of $(q,I)$ for preparation $j$, with the covariance defined using $1/N$, then
	\[
	\overline E_j(t)
	=
	\underbrace{\frac{1}{2}\mu_j^T G\mu_j}_{E_{\mu,j}(t)}
	+
	\underbrace{\frac{1}{2}\operatorname{Tr}[G\Sigma_j]}_{E_{\Sigma,j}(t)}.
	\]
	The first contribution is the energy associated with the center of the ensemble in phase space; the second measures the energy contained in its spread. This decomposition provides a direct bridge between the measured noisy trajectories and the Gaussian description developed in Sec.~\ref{sec:thermal_ou}, without interpreting the laboratory drive itself as an equilibrium thermal reservoir.
	
	On the trigger-time coordinate of the noisy records, the two ensemble-mean energies cross at $t_\times^{\rm noise}=\SI{1.55}{ms}$. Resampling complete trajectories within each preparation gives a 95\% bootstrap interval of $\SIrange{1.46810}{1.62066}{ms}$ from 50\,000 resamples. At late times, between 40 and $\SI{100}{ms}$, the covariance term accounts for approximately $90.2\%$ of the fast ensemble energy and $87.6\%$ of the slow ensemble energy. Figure~\ref{fig:stochastic_data}(a) shows the persistence of the ordering inversion together with the run-to-run spread and the bootstrap interval around the crossing, while Fig.~\ref{fig:stochastic_data}(b) shows how the covariance contribution grows to dominate the late-time ensemble energy. The engineered stochastic drive therefore preserves the preparation-controlled crossing at the ensemble level while transferring most of the surviving late-time energy into fluctuations about the ensemble mean.
	
	The two measurements establish complementary aspects of the same preparation-controlled relaxation. In the deterministic traces, the higher-energy fast preparation crosses below the lower-energy slow preparation on the millisecond scale. With the engineered noise drive present, the crossing remains visible in the ensemble-averaged energy, while fluctuations make an increasingly important contribution at late times. The laboratory result is therefore a direct test of the modal mechanism developed in Sec.~\ref{sec:deterministic_modes}; the equilibrium thermal problem considered next is a separate theoretical extension in which the stochastic forcing is tied to dissipation by the fluctuation--dissipation relation.
	
	\section{Thermal fluctuations and Gaussian relaxation in the Ornstein--Uhlenbeck circuit}
	\label{sec:thermal_ou}
	
	\begin{figure}[htbp]
		\centering
		\includegraphics[width=\textwidth]{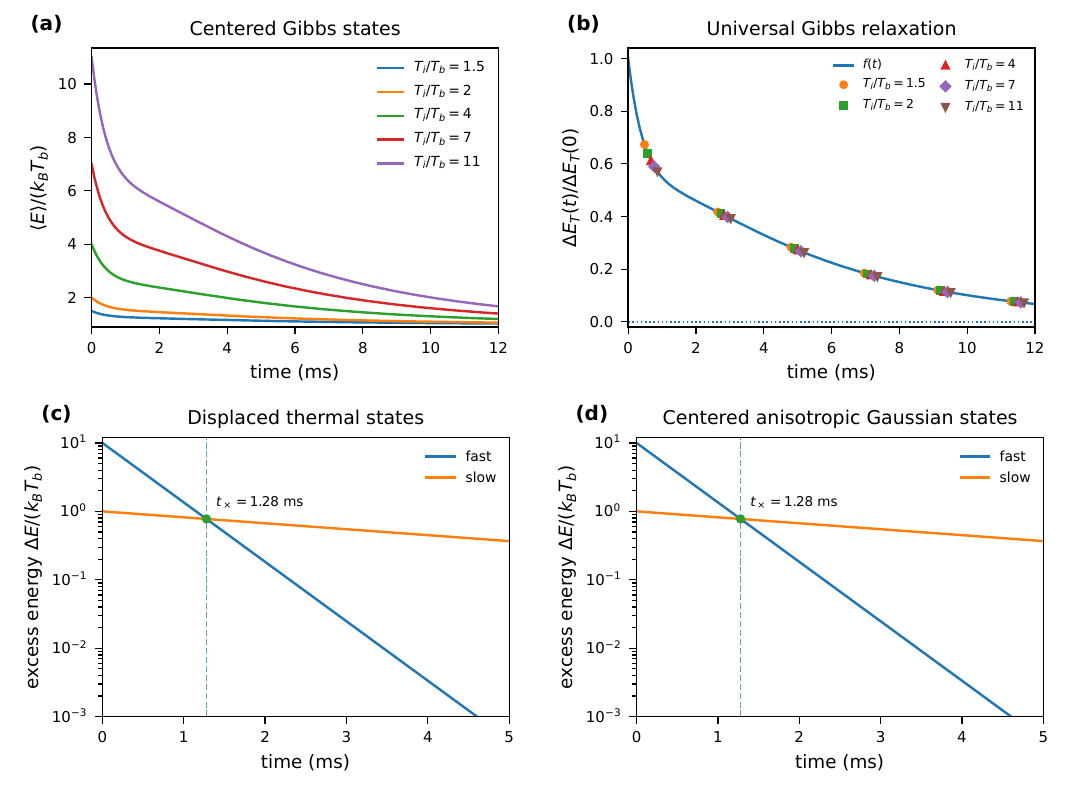}
		\caption{Thermal ordering and nonequilibrium crossings in the Ornstein--Uhlenbeck circuit. (a) Mean energy for several centered Gibbs preparations with different initial temperatures $T_i$. Their ordering is preserved throughout the relaxation. (b) After subtracting the common bath energy and normalizing by the initial excess, all centered Gibbs preparations collapse onto the same positive relaxation function $\Delta E_T(t)/\Delta E_T(0)=f(t)$. This universal collapse shows that the absence of a crossing is a property of the centered Gibbs family, rather than a consequence of a particular temperature choice. (c) For displaced thermal states, the nonequilibrium information is carried by the first moment: a higher-excess-energy preparation aligned with the fast mode crosses below a lower-excess-energy slow preparation. (d) For centered anisotropic Gaussian states, the mean vanishes and the same ordering inversion is carried by the covariance. Panels (c) and (d) show the excess energy $\Delta E=\langle E\rangle-\langle E\rangle_b$ on a logarithmic scale; subtracting the common bath contribution does not change the crossing condition.}
		\label{fig:thermal_nonequilibrium_result}
	\end{figure}
	
	The deterministic discussion treated the circuit state as a single point moving through the $(q,I)$ plane.  At nonzero temperature, that picture must be enlarged: the resistor that removes energy from the circuit also produces voltage fluctuations.  Instead of following one trajectory, we therefore follow a distribution of possible circuit states.  In the classical white-noise regime, the thermal voltage of a resistor is described by the Johnson--Nyquist relation \cite{Johnson1928,Nyquist1928}, with pedagogical discussions of thermal-noise measurements in Refs.~\cite{Abbott1996ThermalNoise,Mishonov2022JohnsonNoise,Pruttivarasin2018Noise,Peng2026Noise}.  For a series RLC circuit coupled to a bath at temperature $T_b$,
	\begin{equation}
		L\ddot q+R\dot q+\frac{q}{C}=\xi(t),
		\qquad
		\langle \xi(t)\xi(t')\rangle=2R\kb T_b\,\delta(t-t').
		\label{eq:langevin_rlc}
	\end{equation}
	The random voltage $\xi(t)$ represents the bath fluctuations.\footnote{The coefficient $2Rk_{\rm B}T_b$ is the two-sided time-domain convention.  The familiar $4Rk_{\rm B}T_b$ Johnson--Nyquist expression is the corresponding one-sided voltage-noise spectral density.  The white-noise approximation also assumes that the frequencies relevant to the circuit lie well inside the classical, effectively flat part of the resistor-noise spectrum.}  This Langevin description is a standard bridge between fluctuating electrical circuits and Brownian motion \cite{Vasziova2010,Tothova2011,NorrelykkeFlyvbjerg2011}; broader treatments of stochastic processes and the Ornstein--Uhlenbeck model can be found in Refs.~\cite{Gardiner2009StochasticMethods,Risken1989FokkerPlanck}, while noisy RLC networks provide a useful setting for stochastic thermodynamics \cite{Freitas2020RLC}.
	
	Using the same state vector $X=(q,I)^T$ as in Sec.~\ref{sec:deterministic_modes}, the fluctuating circuit can be written as
	\begin{equation}
		\dd X=A X\,\dd t+B\,\dd W_t,
		\qquad
		B=\begin{pmatrix}0\\ \sqrt{2R\kb T_b}/L\end{pmatrix},
		\label{eq:ito_rlc}
	\end{equation}
	where $W_t$ is a standard Wiener process.  The deterministic drift matrix $A$ is unchanged: the noise does not replace the slow and fast modes found earlier, but continually perturbs the circuit while those modes relax.  We use the It\^o convention throughout; because the noise is additive, the equivalent Stratonovich equation has the same drift.\footnote{For readers new to stochastic calculus, the main point needed here is that $\dd W_t$ scales as $\sqrt{\dd t}$ rather than as $\dd t$.  Consequently, when a nonlinear function such as the energy is differentiated, terms quadratic in $\dd W_t$ survive after averaging.  This is the origin of the thermal-injection term in Eq.~\eqref{eq:stochastic_energy_balance}.}
	
	For an ensemble of circuit states, two quantities have an immediate geometric meaning: the mean
	$\mu(t)=\langle X(t)\rangle$, which gives the center of the cloud in phase space, and the covariance
	$\Sigma(t)=\langle[X(t)-\mu(t)][X(t)-\mu(t)]^T\rangle$, which describes its width and orientation.  They obey
	\begin{equation}
		\dot\mu=A\mu,
		\qquad
		\dot\Sigma=A\Sigma+\Sigma A^T+BB^T.
		\label{eq:mean_cov}
	\end{equation}
	The first equation is exactly the deterministic RLC equation.  Thermal fluctuations therefore leave the motion of the ensemble center unchanged on average.  Their effect appears in the second equation, through the evolution of the cloud shape.  These two moment equations are valid beyond Gaussian statistics; when the state distribution is Gaussian, however, $\mu$ and $\Sigma$ contain all of its information.
	
	The competition between dissipation and thermal fluctuations is especially clear at the level of the energy.  Applying It\^o calculus to
	$E=q^2/(2C)+LI^2/2$ gives
	\begin{equation}
		\frac{\dd}{\dd t}\langle E\rangle
		=-R\left(\langle I^2\rangle-\frac{\kb T_b}{L}\right).
		\label{eq:stochastic_energy_balance}
	\end{equation}
	The term $-R\langle I^2\rangle$ is the Joule loss already encountered in the deterministic circuit.  The second term is the average energy supplied by the bath.  At equilibrium the two balance, giving $\langle I^2\rangle_b=\kb T_b/L$.  The stationary covariance is
	\begin{equation}
		\Sigma_b=\kb T_b\,G^{-1}
		=
		\begin{pmatrix}
			C\kb T_b & 0\\
			0 & \kb T_b/L
		\end{pmatrix},
		\label{eq:gibbs_cov}
	\end{equation}
	so the capacitor and inductor each carry an average energy $\kb T_b/2$, as required by equipartition.
	
	The ensemble energy naturally separates into a contribution from the center of the cloud and a contribution from its spread,
	\begin{equation}
		\langle E(t)\rangle
		=
		\frac{1}{2}\mu(t)^T G\mu(t)
		+\frac{1}{2}\Tr[G\Sigma(t)].
		\label{eq:mean_energy_decomp}
	\end{equation}
	This decomposition will be useful repeatedly.  The first term is the same quadratic energy that the deterministic model assigns to the mean state.  The second has no deterministic analogue: it is energy stored in fluctuations around that mean.
	
	The connection with the circular geometry of Sec.~\ref{sec:deterministic_modes} becomes particularly transparent after measuring charge and current in thermal units.  Define
	\[
	u=\frac{q}{\sqrt{C\kb T_b}},
	\qquad
	v=\frac{I}{\sqrt{\kb T_b/L}}.
	\]
	In these coordinates the bath covariance is the identity matrix and
	$e=E/(\kb T_b)=(u^2+v^2)/2$, with equilibrium mean $\langle e\rangle_b=1$.  A thermal equilibrium state is therefore represented by a circular Gaussian cloud centered at the origin.  The same slow and fast directions found in the deterministic problem remain present, but they can now appear either in the displacement of the cloud or in its shape.
	
	The simplest thermal preparation is a centered Gibbs state.  Suppose the circuit is initially equilibrated at a temperature $T_i$ and then allowed to relax while coupled to a bath at $T_b$.  The initial mean is zero and the covariance is $\Sigma(0)=\kb T_iG^{-1}$.  Its subsequent covariance is
	\[
	\Sigma(t)=\Sigma_b+
	e^{At}\bigl[\Sigma(0)-\Sigma_b\bigr]e^{A^Tt}.
	\]
	Substituting this result into Eq.~\eqref{eq:mean_energy_decomp} gives
	\begin{equation}
		\langle E(t)\rangle
		=\kb T_b+\kb(T_i-T_b)f(t),
		\qquad
		f(t)=\frac{1}{2}
		\left\|G^{1/2}e^{At}G^{-1/2}\right\|_F^2>0.
		\label{eq:centered_gibbs_no_crossing}
	\end{equation}
	The strict positivity of the relaxation function $f(t)$ follows directly from the properties of positive-definite matrices:
	\begin{equation}
		f(t) = \frac{1}{2} \| G^{1/2} e^{At} G^{-1/2} \|_F^2 = \frac{1}{2} \text{Tr} \left( G^{-1/2} e^{A^T t} G e^{At} G^{-1/2} \right).
	\end{equation}
	Because $G$ is symmetric and positive-definite, the matrix inside the trace remains strictly positive-definite for any finite time $t$. Consequently, its singular values are strictly positive, ensuring $f(t) > 0$ for all $t \ge 0$.
	
	The positivity of $f(t)$ has an immediate consequence.  If two centered Gibbs states begin at temperatures $T_A$ and $T_B$, then their energy difference is simply
	$\kb(T_A-T_B)f(t)$.  Its sign cannot change at any finite time.  In this linear RLC model, changing only the initial temperature therefore changes the amount of excess energy but not its modal composition.  Two centered Gibbs preparations cannot produce an energy-order inversion.
	
	A crossing becomes possible once the preparation contains more information than a single temperature.  The most direct example is a displaced thermal state: the covariance is already equal to $\Sigma_b$, but the cloud center is shifted away from the origin.  The mean then follows the deterministic dynamics, $\mu(t)=e^{At}\mu(0)$, while the thermal covariance contributes the same constant background $\kb T_b$ to every preparation.  If the displacement is aligned with mode $j=s,f$, its excess dimensionless energy obeys $\Delta e_j(t)=\Delta e_j(0)e^{2\lambda_j t}$.  The deterministic crossing of Sec.~\ref{sec:deterministic_modes} is therefore reproduced directly in the first moment of the fluctuating ensemble.  For example, choosing $\Delta e_s(0)=1$ and $\Delta e_f(0)=10$ gives
	\begin{equation}
		t_\times^{(\mu)}
		=
		\frac{\ln[\Delta e_s(0)/\Delta e_f(0)]}
		{2(\lambda_f-\lambda_s)}
		=\SI{1.279}{ms}
		\label{eq:displaced_crossing}
	\end{equation}
	for the reference rates used earlier.  The bath adds the same equilibrium energy to both curves, so it does not alter the crossing of their excess energies.
	
	Nonequilibrium information can also be stored in the shape of the cloud rather than in its center.  Consider a centered Gaussian state with $\mu(0)=0$ but with a covariance that is not proportional to $G^{-1}$.  Such a state is anisotropic in the energy geometry and cannot be described by a single temperature.  Sudden parameter changes, filtered fluctuations, unequal noise in the two circuit variables, or deliberately randomized initial conditions can all generate this kind of preparation.  If $\Delta\Sigma=\Sigma-\Sigma_b$, then
	\begin{equation}
		\Delta\Sigma(t)=e^{At}\Delta\Sigma(0)e^{A^Tt}.
		\label{eq:cov_deviation}
	\end{equation}
	In the eigenbasis of $A$, the three independent covariance components decay with the rates $2\lambda_s$, $\lambda_s+\lambda_f$, and $2\lambda_f$.  The shape of the ensemble can therefore carry the same slow/fast spectral structure as the deterministic state itself \cite{GodrecheLuck2019OU,DeAlmeida2022AnisotropicOU}.
	
	A particularly transparent example is obtained in thermal coordinates by adding extra variance along one eigendirection.  Let $\hat r_s$ and $\hat r_f$ be unit right eigenvectors of $A_y$ and choose
	\[
	\Sigma_s(0)=I+a_s\hat r_s\hat r_s^T,
	\qquad
	\Sigma_f(0)=I+a_f\hat r_f\hat r_f^T,
	\qquad a_f>a_s>0.
	\]
	The equilibrium part $I$ remains unchanged, while each rank-one excess is simply stretched by its corresponding decay mode.  Their excess energies are therefore
	\[
	\Delta e_s(t)=\frac{a_s}{2}e^{2\lambda_s t},
	\qquad
	\Delta e_f(t)=\frac{a_f}{2}e^{2\lambda_f t}.
	\]
	Although both clouds remain centered at the origin, their energy ordering can reverse:
	\begin{equation}
		t_\times^{(\Sigma)}
		=
		\frac{\ln(a_s/a_f)}
		{2(\lambda_f-\lambda_s)}.
		\label{eq:covariance_crossing_time}
	\end{equation}
	For $a_s=2$ and $a_f=20$, the initial excess energies are again $1$ and $10$, and the crossing occurs at the same reference value $\SI{1.279}{ms}$ as in the displaced example.  The two constructions have identical excess-energy curves, but the nonequilibrium information is stored in different places: in the mean for the displaced state and in the covariance for the anisotropic centered state.

	Figure~\ref{fig:thermal_nonequilibrium_result} makes the distinction between equilibrium and nonequilibrium preparations explicit.  Panels~(a) and~(b) show the stronger statement behind Eq.~\eqref{eq:centered_gibbs_no_crossing}: changing the initial temperature of a centered Gibbs state rescales the excess energy but leaves its relaxation shape unchanged.  The normalized curves therefore collapse onto one positive function $f(t)$, so the sign of the energy difference between any two centered Gibbs preparations is fixed for every finite time.  The no-crossing result is thus not a special choice of temperatures or decay rates.
	
	Panels~(c) and~(d) show how the restriction is lifted when the initial ensemble contains thermodynamic nonequilibrium structure beyond temperature alone.  A displacement stores slow or fast spectral content in the first moment, while an anisotropic covariance stores the same information in the second moment even when the mean remains at equilibrium.  For the matched examples shown here, the two nonequilibrium constructions have identical excess-energy curves and the same crossing time, but their physical content is different: one crossing is mean-controlled and the other covariance-controlled.  The drift matrix, bath temperature, and energy observable are unchanged across all four panels; what changes is the structure of the initial preparation.
	
	It is equally important to distinguish the preparation of an ensemble from the mechanism that drives its subsequent fluctuations.  A thermal OU bath continuously injects noise while the covariance relaxes toward $\Sigma_b$.  A randomized initial ensemble followed by deterministic evolution has no continuing stochastic drive, so its covariance simply contracts as $e^{At}\Sigma(0)e^{A^Tt}$.  The engineered laboratory noise of Sec.~\ref{sec:experiment} is a third situation: its effective diffusion is set by the generator, amplifier, coupling coil, and circuit response.  It can be analyzed with the same mean--covariance language, but it becomes equivalent to a Johnson--Nyquist bath only if the noise reaching the RLC circuit is independently calibrated to satisfy the appropriate fluctuation--dissipation relation.
	%%%%%%%%%%%%%%%%%%%%%%%%%%%%%%%%%%%%%%%%%%
	\section{Classical and quantum relatives of the spectral mechanism}
	\label{sec:relatives}
	
	This section develops the spectral picture at a more advanced level,
	moving from the stochastic circuit description to classical and quantum
	generator descriptions. It is intended primarily for beginning graduate
	students with some familiarity with density operators, Markovian master
	equations, and eigenmode methods; useful introductions to the required
	open-system and phase-space methods can be found in
	Refs.~\cite{Orszag2024QuantumOptics,Carmichael1999StatisticalMethods,
		BreuerPetruccione2002}.
	
	We first express the stochastic circuit dynamics in terms of its
	Fokker--Planck generator and then examine Gaussian quantum dynamics
	through Lindblad and Wigner representations. These examples make it
	possible to compare drift, diffusion, preparation, and observability
	within a common linear framework while keeping the classical and
	quantum states physically distinct. We then formulate the same idea
	directly in Liouville space and close with a quantum Brownian example
	whose first-moment drift reproduces the two-mode overdamped structure
	of the classical RLC circuit.
	
	The main text develops only the results needed for these comparisons;
	the companion notebooks provide the longer intermediate derivations,
	phase-space calculations, and numerical checks. The organizing question
	remains the one already encountered in the circuit: which decay modes
	are populated by the initial preparation, and which of those modes are
	visible in the observable used to monitor relaxation?
	
	Before making these broader extensions, it is useful to note an exact
	classical counterpart. The source-free series RLC equation
	$L\ddot q+R\dot q+q/C=0$ has the same mathematical form as the damped
	mechanical oscillator
	\begin{equation}
		m\ddot x+\eta\dot x+kx=0.
	\end{equation}
	The correspondence is
	\begin{equation}
		m\leftrightarrow L,
		\qquad
		\eta\leftrightarrow R,
		\qquad
		k\leftrightarrow \frac{1}{C},
		\qquad
		x\leftrightarrow q,
		\qquad
		\dot x\leftrightarrow I.
	\end{equation}
	The mapping also extends to the stored energy:
	\[
	E_{\rm m}
	=
	\frac{kx^2}{2}
	+
	\frac{m\dot x^2}{2}
	\quad\longleftrightarrow\quad
	E
	=
	\frac{q^2}{2C}
	+
	\frac{LI^2}{2}.
	\]
	In the overdamped regime, the mechanical oscillator therefore has the
	same slow and fast eigendirections and the same preparation-controlled
	ordering mechanism as the deterministic RLC circuit. This is an exact
	correspondence between the two linear equations and their quadratic
	energies, rather than an analogy based only on similar decay curves.
	The stochastic and quantum constructions that follow are different in
	character: they preserve the spectral questions while changing both the
	state space and the generator that acts on it.

	\subsection{From the circuit generator to classical Fokker--Planck dynamics}
	\label{sec:fp_bridge}
	
	The deterministic circuit dynamics introduced in
	Sec.~\ref{sec:deterministic_modes} are generated by the matrix $A$:
	its eigenvalues determine the available decay rates, while the initial
	state determines the amplitudes with which the corresponding modes are
	excited. When thermal noise is included, as in
	Sec.~\ref{sec:thermal_ou}, a single trajectory no longer provides a
	complete description of the system. Repeating the same preparation does
	not produce exactly the same trajectory: one realization may lie at one
	point $X$ in phase space at time $t$, while another realization of the
	same experiment lies somewhere else.
	
	There are then two equivalent ways to describe the stochastic dynamics.
	The Langevin equation follows one random realization at a time. The
	Fokker--Planck description instead follows the complete ensemble by
	assigning a probability density $P(X,t)$ to phase space. The quantity
	$P(X,t)\,dX$ is the probability of finding the system in a small region
	$dX$ around $X$ at time $t$, with normalization
	$\int P(X,t)\,dX=1$. The Fokker--Planck equation is the deterministic
	differential equation that describes how this probability density moves
	and changes shape with time.
	
	For the linear Ornstein--Uhlenbeck dynamics introduced in
	Sec.~\ref{sec:thermal_ou}, deterministic transport and stochastic
	spreading combine to give
	\cite{Gardiner2009StochasticMethods,Risken1989FokkerPlanck}
	\begin{equation}
		\frac{\partial P}{\partial t}
		=
		-\sum_i
		\frac{\partial}{\partial X_i}
		\left[(AX)_i P\right]
		+
		\frac{1}{2}
		\sum_{ij}
		D_{ij}
		\frac{\partial^2 P}{\partial X_i\,\partial X_j}
		\equiv
		\mathcal{L}_{\mathrm{FP}}P,
		\qquad
		D=BB^T .
		\label{eq:fokker_planck}
	\end{equation}
	The first term is the \emph{drift}: it transports the probability
	distribution according to the same deterministic motion that governs
	the noiseless circuit. The second is the \emph{diffusion}: it broadens
	the distribution because nominally identical realizations receive
	different random kicks. Equation~\eqref{eq:fokker_planck} can therefore
	be read as a probability-conservation equation in phase space.
	Probability is redistributed by drift and diffusion, but it is neither
	created nor destroyed.
	
	The distinction between the two stochastic descriptions is useful.
	A Langevin equation contains a random forcing term and generates a
	different trajectory for each realization. Equation~\eqref{eq:fokker_planck}
	contains no random variable: once the initial probability distribution
	is specified, it determines the evolution of $P(X,t)$ deterministically.
	The two descriptions encode the same stochastic process, but they answer
	different questions. The Langevin picture asks how an individual
	realization moves; the Fokker--Planck picture asks how an ensemble of
	such realizations is distributed over phase space.
	
	Equation~\eqref{eq:fokker_planck} is not an additional physical
	assumption.\footnote{One route from the Langevin equation to the
		Fokker--Planck equation is to apply It\^o's formula to an arbitrary
		smooth function $f(X)$, average over the stochastic realizations, and
		then write the same average as
		$\langle f\rangle=\int f(X)P(X,t)\,dX$. Integration by parts transfers
		derivatives from $f$ to $P$, leaving the drift and diffusion operators
		in Eq.~\eqref{eq:fokker_planck}. An equivalent derivation starts from
		the short-time transition probabilities of the Markov process.
		Historically, the equation grew from the diffusion treatments associated
		with Fokker and Planck and is also known, in its general probabilistic
		form, as the forward Kolmogorov equation.}
	Rather, it is the probability-density representation of the stochastic
	dynamics already introduced in Sec.~\ref{sec:thermal_ou}. This also
	explains why the deterministic matrix $A$ has not disappeared: it
	remains inside the larger generator $\mathcal{L}_{\mathrm{FP}}$ as the
	part responsible for drift.
	
	The connection with the mean and covariance used in
	Sec.~\ref{sec:thermal_ou} follows directly. Taking the first moment of
	the probability distribution gives
	\[
	\dot{\mu}=A\mu,
	\]
	so the center of the distribution follows the same slow and fast modes
	as the deterministic circuit. Noise also changes the width and shape of
	the distribution, which are described by its covariance. After
	subtracting the stationary covariance,
	$\Delta\Sigma=\Sigma-\Sigma_b$, the remaining covariance dynamics obey
	\begin{equation}
		\dot{\Delta\Sigma}
		=
		A\Delta\Sigma
		+
		\Delta\Sigma A^T .
		\label{eq:sec5_cov_generator}
	\end{equation}
	
	This equation reveals spectral structure that is absent when only a
	single deterministic trajectory is considered. In the slow--fast
	eigenbasis of $A$, with deterministic eigenvalues $\lambda_s$ and
	$\lambda_f$, the independent covariance components satisfy
	\begin{equation}
		\dot{\sigma}_{ss}
		=
		2\lambda_s\sigma_{ss},
		\qquad
		\dot{\sigma}_{sf}
		=
		(\lambda_s+\lambda_f)\sigma_{sf},
		\qquad
		\dot{\sigma}_{ff}
		=
		2\lambda_f\sigma_{ff}.
		\label{eq:sec5_cov_rates}
	\end{equation}
	Their decay exponents are therefore built from pairwise sums of the
	original deterministic eigenvalues. A fluctuation entirely within the
	slow direction evolves with $2\lambda_s$, a correlation between the
	slow and fast directions with $\lambda_s+\lambda_f$, and a fluctuation
	entirely within the fast direction with $2\lambda_f$. The first moments
	retain the original slow and fast exponents, whereas the second moments
	expose this additional set of spectral combinations.
	
	The full probability density generally contains more information than
	its first two moments. For the linear Ornstein--Uhlenbeck problem,
	however, a Gaussian initial distribution remains Gaussian under the
	evolution. Its center is specified by $\mu$, its shape by $\Sigma$, and
	these quantities therefore provide a complete description within the
	Gaussian family. The companion notebook derives
	Eq.~\eqref{eq:fokker_planck} from the stochastic equation, obtains the
	moment equations explicitly, and verifies the predicted decay rates
	numerically.
	
	The usefulness of the Fokker--Planck formulation becomes particularly
	clear in the quantum problem. A Markovian quantum harmonic oscillator
	can also be represented in phase space, where its Lindblad master
	equation becomes a Fokker--Planck-type equation for a quasiprobability
	distribution. Drift, diffusion, and spectral decay can then be compared
	directly with the stochastic circuit, although a classical probability
	density and a quantum quasiprobability do not have the same physical
	interpretation.

	\subsection{From Lindblad dynamics to a Wigner Fokker--Planck equation}
	\label{sec:qho_phase_space}
	
	For a Markovian open quantum system, the state is described by a density
	operator $\rho$. Its evolution can be written in
	Gorini--Kossakowski--Sudarshan--Lindblad form
	\cite{Gorini1976GKSL,Lindblad1976,Manzano2020Lindblad,Campaioli2024QME},
	\begin{equation}
		\dot{\rho}
		=
		-\frac{i}{\hbar}[H,\rho]
		+
		\sum_j \gamma_j
		\left(
		L_j\rho L_j^\dagger
		-
		\frac{1}{2}\{L_j^\dagger L_j,\rho\}
		\right)
		\equiv
		\mathcal{L}\rho .
		\label{eq:lindblad}
	\end{equation}
	The first term gives the unitary evolution generated by the Hamiltonian
	$H$. The remaining terms describe coupling to the environment: each
	operator $L_j$ specifies a dissipative channel and $\gamma_j$ its rate,
	while $\{A,B\}=AB+BA$ denotes the anticommutator. The superoperator
	$\mathcal L$ therefore plays, for the density operator, a role analogous
	to that played by the dynamical generators introduced earlier for the
	circuit.
	
	For a harmonic oscillator, this connection can be made particularly
	transparent by representing the quantum state in phase space. Instead
	of working directly with the operator $\rho$, one associates with it a
	function of the oscillator quadratures. Several such representations
	are possible, including the Glauber--Sudarshan $P$, Husimi $Q$, and
	Wigner functions. We use the Wigner function $W(x,p,t)$ because its
	first moments and symmetrized second moments coincide directly with the
	quadrature moments used below
	\cite{Orszag2024QuantumOptics,Carmichael1999StatisticalMethods}.
	
	We define dimensionless quadratures
	\begin{equation}
		x=\frac{a+a^\dagger}{\sqrt{2}},
		\qquad
		p=\frac{a-a^\dagger}{i\sqrt{2}},
		\qquad
		[x,p]=i ,
		\label{eq:qho_quadratures}
	\end{equation}
	and collect them in $R=(x,p)^T$. Their mean and covariance are
	\begin{equation}
		\mu=\langle R\rangle,
		\qquad
		V_{ij}
		=
		\frac{1}{2}
		\left\langle
		\Delta R_i\Delta R_j
		+
		\Delta R_j\Delta R_i
		\right\rangle .
		\label{eq:qho_moments}
	\end{equation}
	
	For the zero-affine-drift case used below---quadratic Hamiltonians and
	linear Lindblad operators after any constant phase-space drift has been
	removed by an appropriate displacement---the operator master equation
	can be transformed into a drift--diffusion equation for the Wigner
	function,
	\begin{equation}
		\frac{\partial W}{\partial t}
		=
		-\sum_i
		\frac{\partial}{\partial R_i}
		\left[(A_qR)_iW\right]
		+
		\frac{1}{2}
		\sum_{ij}
		(D_q)_{ij}
		\frac{\partial^2W}
		{\partial R_i\,\partial R_j}
		\equiv
		\mathcal{L}_W W .
		\label{eq:wigner_fp_general}
	\end{equation}
	Here $A_q$ is the quantum drift matrix and $D_q$ the diffusion matrix
	determined by the Hamiltonian and dissipative channels. The detailed
	operator-to-phase-space transformation is given in the companion
	notebook. What matters for the present discussion is that
	Eq.~\eqref{eq:wigner_fp_general} has the same mathematical
	drift--diffusion structure as the classical Fokker--Planck equation in
	Eq.~\eqref{eq:fokker_planck}. Consequently,
	\begin{equation}
		\dot{\mu}=A_q\mu,
		\qquad
		\dot V=A_qV+VA_q^T+D_q ,
		\label{eq:qho_general_moments}
	\end{equation}
	which has exactly the same matrix form as the classical
	Ornstein--Uhlenbeck moment equations.
	
	The similarity is mathematical rather than an identification of the two
	states. A classical probability density is nonnegative and assigns
	probabilities directly to regions of phase space. A Wigner function is
	a quasiprobability representation of a quantum state and can take
	negative values. Quantum covariances are also constrained by the
	uncertainty relation
	\begin{equation}
		V+\frac{i}{2}\Omega\geq0,
		\qquad
		\Omega=
		\begin{pmatrix}
			0&1\\
			-1&0
		\end{pmatrix},
		\label{eq:qho_uncertainty}
	\end{equation}
	a restriction with no classical counterpart.
	
	The usefulness of the correspondence is therefore precise. Within this
	Gaussian class, the classical and quantum problems share the same
	linear algebra for drift, diffusion, means, and covariances, while the
	objects being evolved and the allowed states remain physically
	different. We can now use this common structure to ask whether the
	slow--fast spectral mechanism responsible for the circuit crossing also
	appears in a thermally damped quantum harmonic oscillator.

	\subsection{Thermally damped quantum oscillator: one mean-energy decay law}
	\label{sec:qho_thermal}
	
	We begin with the standard harmonic oscillator coupled to a thermal
	Markov bath. This is the simplest quantum model in which to test whether
	the classical preparation mechanism survives after the state has been
	promoted from a probability distribution to a density operator. Its
	master equation is
	\begin{equation}
		\dot\rho
		=
		-i[\omega a^\dagger a,\rho]
		+
		\kappa(\bar n+1)\mathcal D[a]\rho
		+
		\kappa\bar n\,\mathcal D[a^\dagger]\rho,
		\qquad
		\mathcal D[L]\rho
		=
		L\rho L^\dagger-\frac12\{L^\dagger L,\rho\},
		\label{eq:qho_master_main}
	\end{equation}
	where
	\[
	\bar n
	=
	\left[
	\exp\left(\frac{\hbar\omega}{\kb T_b}\right)-1
	\right]^{-1}
	\]
	is the thermal occupation of the bath and $\kappa$ is the damping rate.
	The first dissipator describes loss of oscillator quanta into the bath,
	while the second describes thermally induced excitation. Together they
	drive the oscillator toward the Gibbs state at temperature $T_b$.
	
	Using the Wigner representation introduced above,
	Eq.~\eqref{eq:qho_master_main} becomes
	\cite{Orszag2024QuantumOptics,Carmichael1999StatisticalMethods}
	\begin{align}
		\frac{\partial W}{\partial t}
		={}&
		-\frac{\partial}{\partial x}
		\left[
		\left(-\frac{\kappa}{2}x+\omega p\right)W
		\right]
		-
		\frac{\partial}{\partial p}
		\left[
		\left(-\omega x-\frac{\kappa}{2}p\right)W
		\right]
		\notag\\
		&+
		\frac{\kappa\nu_b}{2}
		\left(
		\frac{\partial^2W}{\partial x^2}
		+
		\frac{\partial^2W}{\partial p^2}
		\right),
		\qquad
		\nu_b=\bar n+\frac12 .
		\label{eq:qho_thermal_wigner_fp}
	\end{align}
	The structure is easy to read physically. The Hamiltonian rotates the
	distribution in the $(x,p)$ plane, damping contracts both quadratures at
	the same rate $\kappa/2$, and the bath adds isotropic diffusion. In the
	matrix notation of the preceding subsections,
	\begin{align*}
		A_0&=-\frac{\kappa}{2}I+\omega J,
		&
		J&=
		\begin{pmatrix}
			0&1\\
			-1&0
		\end{pmatrix},
		&
		D_0&=\kappa\nu_b I,
		\\
		\dot\mu&=A_0\mu,
		&
		\dot V&=A_0V+VA_0^T+D_0 .
	\end{align*}
	The stationary covariance is therefore $V_b=\nu_b I$. Because both the
	damping and diffusion are rotationally symmetric, the solutions separate
	into an overall decay and a phase-space rotation,
	\begin{align}
		\mu(t)
		&=
		e^{-\kappa t/2}
		\mathcal R(\omega t)\mu(0),
		\label{eq:qho_mu_solution}\\
		V(t)-V_b
		&=
		e^{-\kappa t}
		\mathcal R(\omega t)
		[V(0)-V_b]
		\mathcal R^T(\omega t),
		\label{eq:qho_V_solution}
	\end{align}
	where $\mathcal R(\omega t)$ is the rotation matrix generated by $J$.
	
	These equations provide quantum counterparts of the two stochastic
	preparations discussed in Sec.~\ref{sec:thermal_ou}. A displaced thermal
	state changes the first moment while leaving the covariance equal to
	$V_b$. A centered squeezed thermal state leaves the first moment at the
	origin but changes the covariance, paralleling the centered anisotropic
	Gaussian ensemble. The analogy concerns their first- and second-moment
	structure; the quantum covariance must still satisfy the uncertainty
	relation discussed above.
	
	At this point the comparison with the RLC circuit becomes instructive
	because the analogy fails in a precise way. For the bare oscillator
	Hamiltonian,
	\begin{align*}
		H_0
		&=
		\hbar\omega\left(a^\dagger a+\frac12\right)
		=
		\frac{\hbar\omega}{2}(x^2+p^2),
		\\
		\langle H_0\rangle
		&=
		\frac{\hbar\omega}{2}
		\left(
		\mu^T\mu+\Tr V
		\right).
	\end{align*}
	The energy depends only on the rotationally invariant quantities
	$\mu^T\mu$ and $\Tr V$. Since the rotations in
	Eqs.~\eqref{eq:qho_mu_solution} and \eqref{eq:qho_V_solution} do not
	change either quantity, the excess mean energy above the bath value
	satisfies
	\begin{equation}
		\Delta E(t)
		\equiv
		\langle H_0(t)\rangle-\langle H_0\rangle_b
		=
		e^{-\kappa t}\Delta E(0).
		\label{eq:qho_no_energy_crossing}
	\end{equation}
	Thus displaced and squeezed preparations can carry different
	nonequilibrium information, but the mean bare-energy observable sees
	only one decay envelope. For any two such preparations,
	\begin{equation}
		E_1(t)-E_2(t)
		=
		e^{-\kappa t}
		[E_1(0)-E_2(0)],
		\label{eq:qho_energy_ordering}
	\end{equation}
	so their mean-energy ordering cannot reverse. 
	
	The result is stronger than the Gaussian phase-space argument used to
	obtain it. Acting with the adjoint dissipator on the number operator
	$n=a^\dagger a$ gives
	\begin{align}
		\mathcal D^\dagger[a]n&=-n,
		&
		\mathcal D^\dagger[a^\dagger]n&=n+1,
		\notag\\
		\frac{\dd}{\dd t}\langle n\rangle
		&=
		-\kappa(\langle n\rangle-\bar n).
		\label{eq:qho_n_general}
	\end{align}
	Hence Eq.~\eqref{eq:qho_energy_ordering} holds for any initial state
	with finite mean occupation, not only for Gaussian states. This is a
	no-crossing result for the \emph{mean bare energy} under the
	phase-insensitive thermal generator; it does not exclude quantum Mpemba
	behavior defined through populations, coherences, state-space distances,
	or other observables
	\cite{Longhi2025QHO,Kheirandish2025QHO,Ares2025Review}.
	
	The contrast with the overdamped RLC circuit is now clear. In the
	circuit, the generator supplies two real decay directions that the
	energy can weight differently. In the ordinary thermal quantum
	oscillator, both quadratures have the same damping envelope and the bare
	energy is rotationally symmetric, so directional information carried by
	displacement or squeezing cannot change the ordering of the mean energy.
	To recover an energy crossing, the generator itself must provide
	distinct relaxation sectors that are visible to that observable.

	\subsection{An anisotropic Gaussian oscillator: restoring slow and fast sectors}
	\label{sec:qho_parametric}
	
	The preceding negative result identifies what is missing: the ordinary
	thermal oscillator has no slow--fast anisotropy in its quadrature
	damping. A minimal way to introduce one is a resonantly driven
	degenerate parametric oscillator. In a frame rotating at the bare
	oscillator frequency and within the usual rotating-wave description, we
	take \cite{CollettGardiner1984,Orszag2024QuantumOptics}
	\begin{equation}
		H_\epsilon
		=
		\frac{i\hbar\epsilon}{2}
		\left(a^{\dagger 2}-a^2\right),
		\label{eq:parametric_H}
	\end{equation}
	with the same thermal dissipator as in
	Eq.~\eqref{eq:qho_master_main}. The pump phase can be absorbed into the
	definition of the rotating quadratures, so $\epsilon>0$ may be chosen
	without loss of generality. The parametric interaction amplifies one
	quadrature while deamplifying the other. Including thermal damping,
	their first moments obey
	\begin{equation}
		\dot x
		=
		\left(-\frac{\kappa}{2}+\epsilon\right)x,
		\qquad
		\dot p
		=
		\left(-\frac{\kappa}{2}-\epsilon\right)p.
		\label{eq:parametric_quadrature_drift}
	\end{equation}
	The corresponding Wigner equation is
	\begin{align}
		\frac{\partial W}{\partial t}
		={}&
		-\frac{\partial}{\partial x}
		\left[
		\left(-\frac{\kappa}{2}+\epsilon\right)xW
		\right]
		-
		\frac{\partial}{\partial p}
		\left[
		\left(-\frac{\kappa}{2}-\epsilon\right)pW
		\right]
		\notag\\
		&+
		\frac{\kappa\nu_b}{2}
		\left(
		\frac{\partial^2W}{\partial x^2}
		+
		\frac{\partial^2W}{\partial p^2}
		\right),
		\label{eq:parametric_wigner_fp}
	\end{align}
	so that
	\begin{equation}
		A_\epsilon
		=
		\begin{pmatrix}
			-\kappa/2+\epsilon&0\\
			0&-\kappa/2-\epsilon
		\end{pmatrix},
		\qquad
		D_\epsilon=\kappa\nu_b I .
		\label{eq:parametric_AD}
	\end{equation}
	Unlike the thermal oscillator of the previous subsection, the drift is
	now anisotropic while the bath diffusion remains isotropic. The
	quadratures themselves are the slow and fast eigendirections,
	\begin{equation}
		\lambda_s=-\frac{\kappa}{2}+\epsilon,
		\qquad
		\lambda_f=-\frac{\kappa}{2}-\epsilon,
		\qquad
		0<\epsilon<\frac{\kappa}{2}.
		\label{eq:parametric_rates}
	\end{equation}
	The last inequality is the below-threshold condition: both eigenvalues
	remain negative, so the driven oscillator relaxes to a stationary
	Gaussian state rather than becoming unstable.
	
	For the diagonal drift in Eq.~\eqref{eq:parametric_AD}, the stationary
	Lyapunov equation can be solved component by component,
	\begin{align}
		2\lambda_s(V_{\rm ss})_{xx}
		+\kappa\nu_b
		&=0,
		&
		2\lambda_f(V_{\rm ss})_{pp}
		+\kappa\nu_b
		&=0,
		\notag\\
		(\lambda_s+\lambda_f)(V_{\rm ss})_{xp}
		&=0,
		\notag\\
		V_{\rm ss}
		&=
		\nu_b
		\begin{pmatrix}
			\displaystyle\frac{\kappa}{\kappa-2\epsilon}&0\\[7pt]
			0&\displaystyle\frac{\kappa}{\kappa+2\epsilon}
		\end{pmatrix}.
		\label{eq:parametric_stationary_cov}
	\end{align}
	The stationary state is physical throughout the stable regime because
	\begin{equation}
		\det V_{\rm ss}
		=
		\nu_b^2
		\frac{\kappa^2}{\kappa^2-4\epsilon^2}
		\geq
		\nu_b^2
		\geq
		\frac14 .
	\end{equation}
	It is not, however, a Gibbs state of the bare Hamiltonian $H_0$: the
	parametric pump is continuously present and maintains a driven
	stationary Gaussian state. The comparison below therefore concerns
	relaxation toward that stationary state, not thermalization. The number
	operator is unchanged by the rotation used to define the interaction
	frame, so the bare energy
	$H_0=\hbar\omega(n+1/2)$ remains a well-defined measure of the
	laboratory-mode occupation. We measure its excess relative to the
	driven stationary state,
	\[
	\Delta E(t)
	=
	\langle H_0\rangle_t
	-
	\langle H_0\rangle_{\rm ss}.
	\]
	
	We first place the nonequilibrium information in the first moment while
	keeping the covariance fixed at $V_{\rm ss}$. A displacement prepared
	purely along the slow or fast quadrature evolves as
	\begin{align*}
		\mu_s(t)
		&=
		\mu_s(0)e^{\lambda_s t},
		&
		\mu_f(t)
		&=
		\mu_f(0)e^{\lambda_f t},
		\\
		\Delta E_{\mu,j}(t)
		&=
		\frac{\hbar\omega}{2}\mu_j^2(t),
		&
		j&\in\{s,f\},
		\\
		\Delta E_{\mu,s}(t)
		&=
		\Delta E_{\mu,s}(0)e^{2\lambda_s t},
		&
		\Delta E_{\mu,f}(t)
		&=
		\Delta E_{\mu,f}(0)e^{2\lambda_f t}.
	\end{align*}
	Because the two preparations share the same covariance, their excess
	bare energies differ only through their displacements. If the fast
	preparation starts with the larger excess energy, the ordering reverses
	at
	\begin{equation}
		t_\times^{(\mu)}
		=
		\frac{
			\ln[
			\Delta E_{\mu,s}(0)/
			\Delta E_{\mu,f}(0)
			]
		}{
			2(\lambda_f-\lambda_s)
		}
		=
		\frac{1}{4\epsilon}
		\ln
		\frac{
			\Delta E_{\mu,f}(0)
		}{
			\Delta E_{\mu,s}(0)
		}.
		\label{eq:parametric_displacement_crossing}
	\end{equation}
	This is the same crossing algebra as for preparations along the slow and
	fast eigendirections of the RLC circuit. The essential change relative
	to the ordinary thermal QHO is not the preparation but the generator:
	the parametric interaction has created two distinct decay rates that the
	bare energy can distinguish.
	
	The same mechanism can be stored entirely in the covariance. Let
	$\Delta V=V-V_{\rm ss}$ and consider centered Gaussian preparations with
	excess covariance placed along one quadrature,
	\begin{equation}
		\Delta V_s(0)
		=
		\delta_s
		\begin{pmatrix}
			1&0\\
			0&0
		\end{pmatrix},
		\qquad
		\Delta V_f(0)
		=
		\delta_f
		\begin{pmatrix}
			0&0\\
			0&1
		\end{pmatrix},
		\qquad
		\delta_s,\delta_f>0.
		\label{eq:parametric_cov_preps}
	\end{equation}
	Because these increments are positive semidefinite, adding them to the
	physical stationary covariance preserves the one-mode uncertainty
	condition. The resulting centered states are physical Gaussian
	preparations whose excess fluctuations are concentrated predominantly
	along one of the two dynamical quadratures. Their covariance deviation
	evolves as
	\begin{align}
		\Delta V(t)
		&=
		e^{A_\epsilon t}
		\Delta V(0)
		e^{A_\epsilon^Tt},
		\label{eq:parametric_cov_evolution}\\
		\Delta E_V(t)
		&=
		\frac{\hbar\omega}{2}\Tr[\Delta V(t)].
		\notag
	\end{align}
	For the two preparations in Eq.~\eqref{eq:parametric_cov_preps},
	\begin{equation}
		\Delta E_{V,s}(t)
		=
		\Delta E_{V,s}(0)e^{2\lambda_s t},
		\qquad
		\Delta E_{V,f}(t)
		=
		\Delta E_{V,f}(0)e^{2\lambda_f t},
	\end{equation}
	so that
	\begin{equation}
		t_\times^{(V)}
		=
		\frac{
			\ln[
			\Delta E_{V,s}(0)/
			\Delta E_{V,f}(0)
			]
		}{
			2(\lambda_f-\lambda_s)
		}.
		\label{eq:parametric_cov_crossing}
	\end{equation}
	The quantum Gaussian problem therefore reproduces the two placements of
	nonequilibrium information already encountered in the classical OU
	circuit: the slow--fast content may reside in the displacement or in
	the covariance.
	
	The covariance dynamics also exposes the role of the observable. Writing
	\begin{align}
		\Delta V
		&=
		\begin{pmatrix}
			v_{ss}&v_{sf}\\
			v_{sf}&v_{ff}
		\end{pmatrix},
		\notag\\
		v_{ss}(t)
		&\propto e^{2\lambda_s t},
		&
		v_{sf}(t)
		&\propto
		e^{(\lambda_s+\lambda_f)t}
		=
		e^{-\kappa t},
		&
		v_{ff}(t)
		&\propto e^{2\lambda_f t},
		\label{eq:parametric_cov_sectors}
	\end{align}
	shows three distinct second-moment sectors. The mixed component $v_{sf}$
	is a genuine dynamical mode, but the bare energy depends on
	$\Tr V=v_{ss}+v_{ff}$ and is therefore blind to it. A mode can thus be
	present in the evolving state without appearing in a particular
	observable.
	
	The correspondence with the RLC circuit is precise but limited. The
	passive circuit dissipates stored energy toward equilibrium, whereas the
	parametric oscillator relaxes toward a nonequilibrium stationary state
	maintained by an external pump. They are not thermodynamically
	equivalent. What they share is the spectral mechanism needed for an
	ordering inversion: distinct decay directions, preparation-dependent
	occupation of those directions, and an observable with nonzero
	sensitivity to the relevant modes. The companion notebook derives the
	Lindblad-to-Wigner mapping in detail, checks the covariance physicality
	conditions, and reproduces
	Eqs.~\eqref{eq:parametric_displacement_crossing} and
	\eqref{eq:parametric_cov_crossing} directly from the master equation.

	\subsection{Liouvillian modes, preparation, and observable selectivity}
	\label{sec:liouvillian_view}
	
	The Gaussian examples make the general structure visible without
	requiring the full machinery of Liouville space. We can now state that
	structure directly. For a finite-dimensional open system, or more
	generally whenever the relevant spectral expansion exists, suppose that
	the Liouvillian $\mathcal L$ has a unique stationary state and is
	diagonalizable. Because $\mathcal L$ need not be Hermitian as a
	superoperator, its right and left eigenoperators form a biorthogonal
	rather than an ordinary orthogonal basis
	\cite{Carollo2021,Ares2025Review}. Using the Hilbert--Schmidt inner
	product
	\[
	\langle\!\langle A|B\rangle\!\rangle
	\equiv
	\Tr(A^\dagger B),
	\]
	we write
	\begin{equation}
		\mathcal L|R_k\rangle\!\rangle
		=
		\Lambda_k|R_k\rangle\!\rangle,
		\qquad
		\langle\!\langle L_k|\mathcal L
		=
		\Lambda_k\langle\!\langle L_k|,
		\qquad
		\langle\!\langle L_j|R_k\rangle\!\rangle
		=
		\delta_{jk}.
	\end{equation}
	The double-ket notation simply treats an operator as a vector in
	operator space. The eigenoperators $R_k$ are spectral directions of the
	evolution; except for the stationary state, they need not themselves be
	physical density operators.
	
	Let $\rho_{\rm ss}$ denote the stationary state and let all
	nonstationary modes satisfy $\operatorname{Re}\Lambda_k<0$. The
	deviation from stationarity can then be expanded as
	\begin{equation}
		|\rho(t)\rangle\!\rangle
		=
		|\rho_{\rm ss}\rangle\!\rangle
		+
		\sum_{k\ge1}
		c_k e^{\Lambda_k t}
		|R_k\rangle\!\rangle,
		\qquad
		c_k
		=
		\langle\!\langle L_k|
		\rho(0)-\rho_{\rm ss}
		\rangle\!\rangle .
		\label{eq:liouvillian_expansion}
	\end{equation}
	This is the operator-space analogue of decomposing the RLC state into
	slow and fast modes. The coefficient $c_k$ measures how strongly the
	initial preparation populates mode $k$. If $c_k=0$, that mode is absent
	from the subsequent relaxation even if it is part of the Liouvillian
	spectrum.
	
	A measured relaxation contains one more selection rule. For a Hermitian
	observable $O$,
	\begin{equation}
		\langle O\rangle_t
		-
		\langle O\rangle_{\rm ss}
		=
		\sum_{k\ge1}
		\underbrace{
			\langle\!\langle L_k|
			\rho(0)-\rho_{\rm ss}
			\rangle\!\rangle
		}_{\text{preparation overlap}}
		\underbrace{
			\langle\!\langle O|R_k\rangle\!\rangle
		}_{\text{observable overlap}}
		e^{\Lambda_k t}.
		\label{eq:observable_spectral_expansion}
	\end{equation}
	A slow Liouvillian mode therefore controls an observed decay only when
	two conditions are met: the preparation must excite it and the
	observable must be sensitive to it. The mixed covariance mode in
	Eq.~\eqref{eq:parametric_cov_sectors} is an explicit example of the
	second condition failing. That mode evolves in the state, but its
	contribution to the bare energy vanishes because the energy depends only
	on the covariance trace. Preparation overlap is central to spectral
	formulations of the quantum Mpemba effect
	\cite{Carollo2021,NavaEgger2024,ZhangQuantumStrong2025,
		Ares2025Review,BeatoTeza2026Relaxation};
	Eq.~\eqref{eq:observable_spectral_expansion} makes equally explicit that
	the relaxation diagnostic selects which populated modes can actually be
	observed.
	
	A dissipative two-level system gives the same distinction in a
	finite-dimensional setting. Write
	\[
	\rho
	=
	\frac12
	(I+x\sigma_x+y\sigma_y+z\sigma_z)
	\]
	and take
	$H=\hbar\omega\sigma_z/2$. A standard Markovian relaxation model gives
	\cite{BreuerPetruccione2002,Campaioli2024QME}
	\begin{equation}
		\dot x=-\omega y-\Gamma_2x,
		\qquad
		\dot y=\omega x-\Gamma_2y,
		\qquad
		\dot z=-\Gamma_1(z-z_{\rm eq}).
	\end{equation}
	The energy depends only on the longitudinal component $z$, while $x$
	and $y$ carry coherence information. In the usual amplitude-damping
	plus pure-dephasing model,
	$\Gamma_2=\Gamma_1/2+\Gamma_\phi$, so the longitudinal and transverse
	decay rates are related but generally different. Two states with the
	same initial $z$ therefore have the same energy and the same
	energy-relaxation curve even when their coherences, and hence their full
	density matrices, relax differently. The lesson is the same as in the
	Gaussian example: a generator can contain dynamical sectors that a
	particular observable does not reveal.
	
	The simple exponential expansion in
	Eq.~\eqref{eq:liouvillian_expansion} assumes diagonalizability. At an
	exceptional point, eigenvalues and eigenvectors coalesce and Jordan
	blocks produce polynomial factors multiplying the exponentials
	\cite{DowningSaroka2025Exceptional}. The modal bookkeeping must then be
	modified, but the organizing questions remain unchanged: which
	dynamical sectors are present, which are excited by the preparation,
	and which are visible to the observable?

	\subsection{Quantum Brownian counterpart}
	\label{sec:CL_counterpart}
	
	The quantum-optical examples above show how drift, diffusion, and
	observable selectivity appear in Gaussian open systems. We close the
	section with a complementary model that makes an even more direct
	connection with the second-order equation of the overdamped RLC
	circuit: quantum Brownian motion described by a
	Caldeira--Leggett-type master equation.\footnote{The
		Caldeira--Leggett approach belongs to the broader theory of quantum
		Brownian motion in which a system coordinate is coupled to a large
		environment that can be represented, in the standard construction, by
		a bath of harmonic oscillators. Its path-integral formulation uses the
		Feynman--Vernon influence functional to eliminate the environmental
		degrees of freedom, leaving both dissipative and fluctuating effects in
		the reduced system dynamics
		\cite{FeynmanVernon1963,CaldeiraLeggett1983}. In the classical limit,
		the same construction recovers Brownian/Fokker--Planck dynamics, making
		the model particularly useful for comparing classical and quantum
		dissipation. For a recent pedagogical and historical account of
		Caldeira's contributions to quantum Brownian motion, see
		Ref.~\cite{BonancaDeffnerIngold2026}. The frequently used local
		Caldeira--Leggett master equations are reduced descriptions obtained
		under additional approximations; unlike the secular quantum-optical
		master equation, such forms are not generically GKSL generators in all
		parameter regimes and can require additional care concerning complete
		positivity and their domain of validity
		\cite{BreuerPetruccione2002}. We therefore use the qualifier
		``Caldeira--Leggett-type'' and use the equation below only for the
		first-moment drift required by the present correspondence.}
	
	The point of this example is deliberately limited. We do not identify
	the full quantum dynamics with the classical circuit. Instead, we show
	that the quantum first moments form a closed linear system whose drift
	has exactly the same mathematical structure as the classical damped
	oscillator.
	
	We retain the dimensionless quadratures introduced in
	Eq.~\eqref{eq:qho_quadratures},
	\[
	x=\frac{a+a^\dagger}{\sqrt{2}},
	\qquad
	p=\frac{a-a^\dagger}{i\sqrt{2}},
	\qquad
	[x,p]=i,
	\]
	and set $\hbar=1$. For
	\begin{equation}
		H_0=\omega_0 a^\dagger a,
		\label{eq:CLHamiltonian}
	\end{equation}
	a convenient Caldeira--Leggett-type form of the dynamics is
	\cite{CaldeiraLeggett1983,de2026collective}
	\begin{align}
		\dot\rho={}&
		-i[H_0,\rho]
		-i\gamma[x,\{p,\rho\}]
		\nonumber\\
		&-
		\gamma
		\coth\left(
		\frac{\omega_0}{2\kb T_b}
		\right)
		[x,[x,\rho]] .
		\label{eq:CLmaster}
	\end{align}
	The friction term acts directly on the momentum quadrature, whereas the
	double commutator describes diffusion. The latter affects second
	moments and fluctuations but not the equations for the first moments.
	This separation is useful here because the RLC correspondence already
	appears completely at the level of the mean quadratures.
	
	For any observable $O$,
	\begin{equation}
		\frac{d}{dt}\langle O\rangle
		=
		\Tr(O\dot\rho).
		\label{eq:CLexpectation}
	\end{equation}
	The Hamiltonian contribution gives
	\begin{equation}
		i[H_0,x]
		=
		\omega_0 p,
		\qquad
		i[H_0,p]
		=
		-\omega_0 x.
		\label{eq:CLHamiltonianCommutators}
	\end{equation}
	The diffusion term does not contribute to either first moment because
	\begin{equation}
		[x,[x,x]]=0,
		\qquad
		[x,[x,p]]=0.
	\end{equation}
	It does, however, enter the second-moment equations and therefore affects
	fluctuations and the mean energy. This parallels the
	mean--covariance separation encountered in the classical
	Ornstein--Uhlenbeck description.
	
	The friction term provides the momentum damping. Using $[p,x]=-i$, its
	contribution gives
	\begin{equation}
		-i\gamma
		\left\langle
		\{[p,x],p\}
		\right\rangle
		=
		-2\gamma\langle p\rangle .
	\end{equation}
	The first moments consequently obey
	\begin{align}
		\frac{d}{dt}\langle x\rangle
		&=
		\omega_0\langle p\rangle,
		\label{eq:CLxmean}\\
		\frac{d}{dt}\langle p\rangle
		&=
		-\omega_0\langle x\rangle
		-
		2\gamma\langle p\rangle .
		\label{eq:CLpmean}
	\end{align}
	Equivalently,
	\begin{equation}
		\frac{d}{dt}
		\begin{pmatrix}
			\langle x\rangle\\
			\langle p\rangle
		\end{pmatrix}
		=
		\underbrace{
			\begin{pmatrix}
				0 & \omega_0\\
				-\omega_0 & -2\gamma
			\end{pmatrix}
		}_{A_{\rm CL}^{(1)}}
		\begin{pmatrix}
			\langle x\rangle\\
			\langle p\rangle
		\end{pmatrix}.
		\label{eq:CLdrift}
	\end{equation}
	
	Eliminating $\langle p\rangle$ gives
	\begin{equation}
		\frac{d^2}{dt^2}\langle x\rangle
		+
		2\gamma
		\frac{d}{dt}\langle x\rangle
		+
		\omega_0^2
		\langle x\rangle
		=
		0.
		\label{eq:CLmean}
	\end{equation}
	The comparison with the source-free RLC circuit,
	\begin{equation}
		\ddot q
		+
		\frac{R}{L}\dot q
		+
		\frac{1}{LC}q
		=
		0,
	\end{equation}
	is therefore immediate:
	\begin{equation}
		q
		\longleftrightarrow
		\langle x\rangle,
		\qquad
		I=\dot q
		\longleftrightarrow
		\omega_0\langle p\rangle,
		\qquad
		\frac{R}{L}
		\longleftrightarrow
		2\gamma,
		\qquad
		\frac{1}{LC}
		\longleftrightarrow
		\omega_0^2.
		\label{eq:RLC_CL_dictionary}
	\end{equation}
	This is a correspondence between the deterministic RLC state and the
	\emph{first moments} of the quantum oscillator; it is not an
	identification of the complete classical and quantum states.
	
	At the level of this drift equation, the motion becomes overdamped when
	$\gamma>\omega_0$. The two real decay rates are then
	\begin{equation}
		\Gamma_{s,f}
		=
		\gamma
		\mp
		\sqrt{\gamma^2-\omega_0^2},
		\qquad
		0<\Gamma_s<\Gamma_f.
		\label{eq:CLrates}
	\end{equation}
	The mean position consequently has the familiar two-mode form
	\begin{equation}
		\langle x(t)\rangle
		=
		C_s e^{-\Gamma_s t}
		+
		C_f e^{-\Gamma_f t}.
		\label{eq:CLsolution}
	\end{equation}
	We emphasize that $\gamma>\omega_0$ is being used here as a statement
	about the first-moment drift matrix. The microscopic validity of a
	particular local Brownian master equation can impose additional
	restrictions on the parameters; those restrictions are separate from
	the algebraic correspondence established here.
	
	The preparation logic encountered in the RLC circuit now follows
	immediately. The slow contribution to the mean motion vanishes when
	\begin{equation}
		\omega_0\langle p(0)\rangle
		=
		-\Gamma_f\langle x(0)\rangle,
		\label{eq:CLfastPreparation}
	\end{equation}
	which is the first-moment counterpart of
	$I(0)=\lambda_f q(0)$ with $\lambda_f=-\Gamma_f$. A suitable initial
	quadrature displacement can therefore suppress the slow contribution
	to the mean motion in the same algebraic way that a suitable choice of
	initial charge and current suppresses the slow circuit mode.
	
	The role of this final example is therefore different from that of the
	parametric oscillator discussed above. The parametric model provides a
	controlled Gaussian setting in which distinct slow and fast sectors
	produce an explicit excess-energy crossing. The Brownian model instead
	shows that the \emph{second-order overdamped equation itself} can
	reappear directly in quantum first-moment dynamics. Together, the two
	examples separate what is structurally common---modal preparation and
	observable selectivity---from what depends on the particular
	open-system generator.
	
	The sequence of examples in this section involves several different
	state spaces and therefore should not be read as a chain of physical
	identifications. Table~\ref{tab:spectral_dictionary} collects the
	correspondences that are actually being used. The deterministic RLC and
	mechanical oscillator share an exact equation-and-energy mapping; the
	classical OU circuit and Gaussian quantum models share a
	drift--diffusion and moment structure; the Caldeira--Leggett example
	shares the overdamped first-moment drift; and the Liouvillian
	description isolates the general preparation--mode--observable
	structure underlying all of these comparisons.
	
	\begin{table}[t]
		\centering
		\footnotesize
		\setlength{\tabcolsep}{4pt}
		\caption{Spectral dictionary for the deterministic, stochastic, and
			quantum descriptions used in Sec.~V. The rows distinguish exact
			dynamical mappings from structural correspondences; they summarize
			the common generator--preparation--observable logic without
			identifying the underlying physical states.}
		\label{tab:spectral_dictionary}
		\begin{tabular}{
				@{}
				p{0.14\textwidth}
				p{0.30\textwidth}
				p{0.23\textwidth}
				p{0.23\textwidth}
				@{}
			}
			\toprule
			\raggedright System
			&
			\raggedright State and generator
			&
			\raggedright Preparation
			&
			\raggedright Observable or diagnostic
			\tabularnewline
			\midrule
			
			\raggedright RLC circuit
			&
			\raggedright
			$X=(q,I)^T$; drift matrix $A$ with slow and fast eigendirections
			&
			\raggedright
			modal amplitudes $c_s,c_f$
			&
			\raggedright
			$q$, $I$, or stored energy $E$
			\tabularnewline
			
			\raggedright Mechanical oscillator
			&
			\raggedright
			$(x,\dot x)^T$; drift matrix obtained by the exact RLC mapping
			&
			\raggedright
			slow and fast modal amplitudes
			&
			\raggedright
			$x$, $\dot x$, or $E_{\rm m}$
			\tabularnewline
			
			\raggedright Classical OU circuit
			&
			\raggedright
			$P(X,t)$ with $\mathcal L_{\rm FP}$; for Gaussian states,
			$(\mu,\Sigma)$ closes
			&
			\raggedright
			displacement and covariance relative to the stationary state
			&
			\raggedright
			moments or mean energy $\langle E\rangle$
			\tabularnewline
			
			\raggedright Thermal QHO
			&
			\raggedright
			$\rho$; phase-insensitive thermal Liouvillian with Wigner
			drift--diffusion representation
			&
			\raggedright
			arbitrary initial state; displacement and squeezing provide
			Gaussian examples
			&
			\raggedright
			mean bare energy has one decay factor; other diagnostics can
			resolve additional state structure
			\tabularnewline
			
			\raggedright Parametric QHO
			&
			\raggedright
			$\rho$ or $W(x,p)$; anisotropic Gaussian drift with rates
			$\lambda_s,\lambda_f$
			&
			\raggedright
			displacement or covariance along slow and fast quadratures
			&
			\raggedright
			excess bare energy or quadrature moments
			\tabularnewline
			
			\raggedright Quantum Brownian oscillator
			&
			\raggedright
			$\rho$; Caldeira--Leggett-type dynamics with first-moment drift
			$A_{\rm CL}^{(1)}$
			&
			\raggedright
			initial quadrature means along slow and fast first-moment
			directions
			&
			\raggedright
			$\langle x\rangle$, $\langle p\rangle$; second moments and
			energy also depend on diffusion
			\tabularnewline
			
			\raggedright General Markovian open system
			&
			\raggedright
			$\rho$; Liouvillian $\mathcal L$ with left and right decay
			eigenoperators when a spectral expansion exists
			&
			\raggedright
			preparation overlaps $c_k$ with dynamical modes
			&
			\raggedright
			observable overlaps
			$\langle\!\langle O|R_k\rangle\!\rangle$
			select visible modes
			\tabularnewline
			
			\bottomrule
		\end{tabular}
	\end{table}
	
	The common object across the rows of
	Table~\ref{tab:spectral_dictionary} is therefore not the physical state
	itself, but the organization of relaxation by a generator, an initial
	preparation, and a chosen diagnostic. The RLC circuit provides the
	simplest experimentally transparent realization of this structure:
	its two decay directions are visible directly in ordinary phase space,
	their weights are controlled by the initial charge and current, and the
	resulting competition can be read out through the stored energy. The
	classical and quantum relatives developed in this section show how the
	same organizing logic persists when the state becomes a probability
	distribution, a covariance matrix, or a density operator, while also
	making clear where the physical equivalence ends.
	%%%%%%%%%%%%%%%%%%%%%%%%%%%%%%%%%%%%%%%
	\section{Conclusion}
	
	An overdamped RLC circuit provides a simple example of why initial energy alone does not determine a relaxation history.  The energy fixes how much is stored in the circuit, while the initial charge and current determine how that energy is distributed between the slow and fast decay modes.  Two states can therefore begin at the same energy and relax differently, or a higher-energy state can fall below a lower-energy state when the former is prepared predominantly along the fast direction.  In energy-normalized phase space, this distinction has a direct geometric meaning: distance from the origin sets the energy, whereas direction controls the modal composition.
	
	The laboratory measurements make this separation operational.  Preparing the circuit close to its slow and fast eigendirections produces the predicted reversal of the energy ordering during deterministic relaxation.  When an externally generated white-noise drive is added, the crossing remains visible in the ensemble-averaged energy, while the late-time signal becomes increasingly associated with fluctuations about the ensemble mean.  This second protocol shows that the preparation-controlled ordering survives the measured engineered stochastic forcing.  It does not by itself establish a thermal realization, which would require an independent calibration connecting the applied noise to the fluctuation--dissipation relation.
	
	By relying on standard benchtop components and direct oscilloscope acquisitions, the minimal experimental setup demonstrates that complex relaxation phenomena, such as modal energy crossings, can be rendered accessible and transparent in an educational setting with accessible resources.
	
	The equilibrium stochastic model clarifies what changes when a single trajectory is replaced by an ensemble.  In the Ornstein--Uhlenbeck description, a centered Gibbs preparation changes only the amount of thermal excitation and preserves the ordering of the mean energy.  Nonequilibrium information can instead be placed in different statistical sectors: a displacement stores slow or fast content in the first moment, while an anisotropic covariance stores it in the shape of a centered ensemble.  In the Fokker--Planck representation these sectors acquire a transparent spectral interpretation.  The first moments retain the deterministic decay rates, whereas the covariance contains the pairwise combinations $2\lambda_s$, $\lambda_s+\lambda_f$, and $2\lambda_f$.
	
	In the present experimental implementation, the stochastic perturbation was injected externally via an auxiliary inductive coupling driven by an arbitrary waveform generator, serving as an engineered noise drive rather than a true thermodynamic reservoir. A natural extension of this work is to implement a calibrated thermal noise source that directly simulates Johnson--Nyquist resistor fluctuations tied to a physical bath temperature $T_b$. By precisely scaling the injected noise intensity to satisfy the fluctuation--dissipation relation, one can experimentally observe the evolution of the system's Gaussian mean--covariance dynamics under genuine thermal fluctuations, enabling a direct laboratory test of thermal ordering, excess-energy crossings, and moment relaxation in fluctuating linear circuits.
	
	The quantum extension shows both the usefulness and the limits of carrying this reasoning to a different physical setting.  For the ordinary thermally damped harmonic oscillator, the Wigner equation has the same drift--diffusion structure as a classical Ornstein--Uhlenbeck process, but the isotropic damping and the rotational symmetry of the bare energy collapse the energy relaxation to a single decay factor.  The adjoint master equation shows that this no-crossing result for the mean bare energy is not restricted to Gaussian preparations: it holds for any initial state with finite mean occupation under the thermal generator considered here.  Introducing a below-threshold parametric drive changes the generator itself.  The two quadratures then acquire distinct slow and fast decay rates, allowing an energy-order inversion when nonequilibrium information is placed either in the displacement or in the covariance.  The mixed covariance sector provides an additional lesson: a dynamical mode may be present in the evolving state while remaining invisible to a particular observable.
	
	The general Liouvillian description makes this structure explicit.  The generator determines the available decay modes, the initial preparation determines their amplitudes, and the observable determines which of those populated modes contribute to the measured relaxation.  The RLC circuit realizes this structure in its simplest form, where both the state space and the energy are directly accessible experimentally.  The stochastic and quantum examples show how the same spectral questions persist when the state is promoted to a probability distribution or a density operator, without requiring the underlying physical systems to be thermodynamically equivalent.
	
	The central lesson is therefore more precise than the statement that some systems relax ``faster'' than others.  A relaxation time is meaningful only after specifying the dynamical generator, the preparation, and the observable used to monitor the approach to the stationary state.  In the circuit, these ingredients can be separated and controlled with elementary linear dynamics and direct measurements.  Their extension to Fokker--Planck and Liouvillian descriptions provides a route from a familiar overdamped transient to the broader spectral organization of nonequilibrium relaxation.
	
	\begin{acknowledgments}
		This study was financed in part by the Coordena\c{c}\~ao de Aperfei\c{c}oamento de Pessoal de N\'ivel Superior -- Brasil (CAPES) -- Finance Code 001. Generative AI tools were used for language editing and literature-search assistance; all scientific content was reviewed and validated by the authors, who take full responsibility for the manuscript.
	\end{acknowledgments}
	
	\section*{Companion calculations and reproducibility}
	Companion notebooks reproduce the numerical calculations,
	experimental-data reconstruction, and manuscript figures, and
	provide editable examples for exploring the deterministic and
	stochastic models. The accompanying documentation specifies the
	execution order, software requirements, and notebook-to-figure map.
	
	\section*{Data availability}

	The complete raw and processed oscilloscope traces, exported analysis data, device specifications, figure-generation scripts, and companion notebooks providing detailed pedagogical derivations of the theories presented in this work are publicly available at \url{https://github.com/mathhenrique-code/Mpemba-RLC}.

	\bibliography{references}
\end{document}